\documentclass[acmlarge, screen]{acmart}

\AtBeginDocument{%
  \providecommand\BibTeX{{%
    \normalfont B\kern-0.5em{\scshape i\kern-0.25em b}\kern-0.8em\TeX}}}

\usepackage{graphicx} % Required for inserting images
\usepackage{pifont}% http://ctan.org/pkg/pifont
\newcommand{\cmark}{\ding{51}}%
\newcommand{\xmark}{\ding{55}}%

\usepackage{subcaption}
\usepackage{multirow}
        
\usepackage{amssymb}
\usepackage{tabularx}
\usepackage{amsmath}
\usepackage{xfrac}
\usepackage{placeins}
\usepackage{xcolor}
\usepackage{color}
\usepackage{array}
\usepackage{colortbl}
\usepackage{enumitem}
\usepackage{xspace}
\usepackage{enumerate}
\usepackage{pifont}
\usepackage{enumitem}
\usepackage{soul}
\usepackage{textcmds}

\setcopyright{none}
\renewcommand\footnotetextcopyrightpermission[1]{} 
\newcommand{\aruna}[1]{{\color{red}{aruna: #1}}}

\newcommand{\prerna}[1]{{\color{blue}{prerna: #1}}}
\newcolumntype{C}{>{\centering\arraybackslash}m{3cm}}

\author{Prerna Khanna}
\email{pkhanna@cs.stonybrook.edu}
\affiliation{%
  \institution{Stony Brook University}
  \city{New York}
  \country{USA}
}

\author{Tanmay Srivastava}
\authornote{Co-first authors with equal contribution}
\email{tsrivastava@cs.stonybrook.edu}
\affiliation{%
  \institution{Stony Brook University}
  \city{New York}
  \country{USA}
}

\author{Shubham Jain}
\email{jain@cs.stonybrook.edu}
\affiliation{%
  \institution{Stony Brook University}
  \city{New York}
  \country{USA}
  }

\author{Aruna Balasubramanian}
\email{arunab@cs.stonybrook.edu}
\affiliation{%
  \institution{Stony Brook University}
  \city{New York}
  \country{USA}
  }

\newcommand{\system}{\emph{UniMotion}\xspace}

\title{\textit{UniMotion}: Self-Supervised Learning for Cross-Domain IMU Motion Recognition}

\begin{document}
\begin{abstract}
%!TEX root = main.tex

\begin{figure}[H]
\centering
\includegraphics[width=0.7\linewidth]{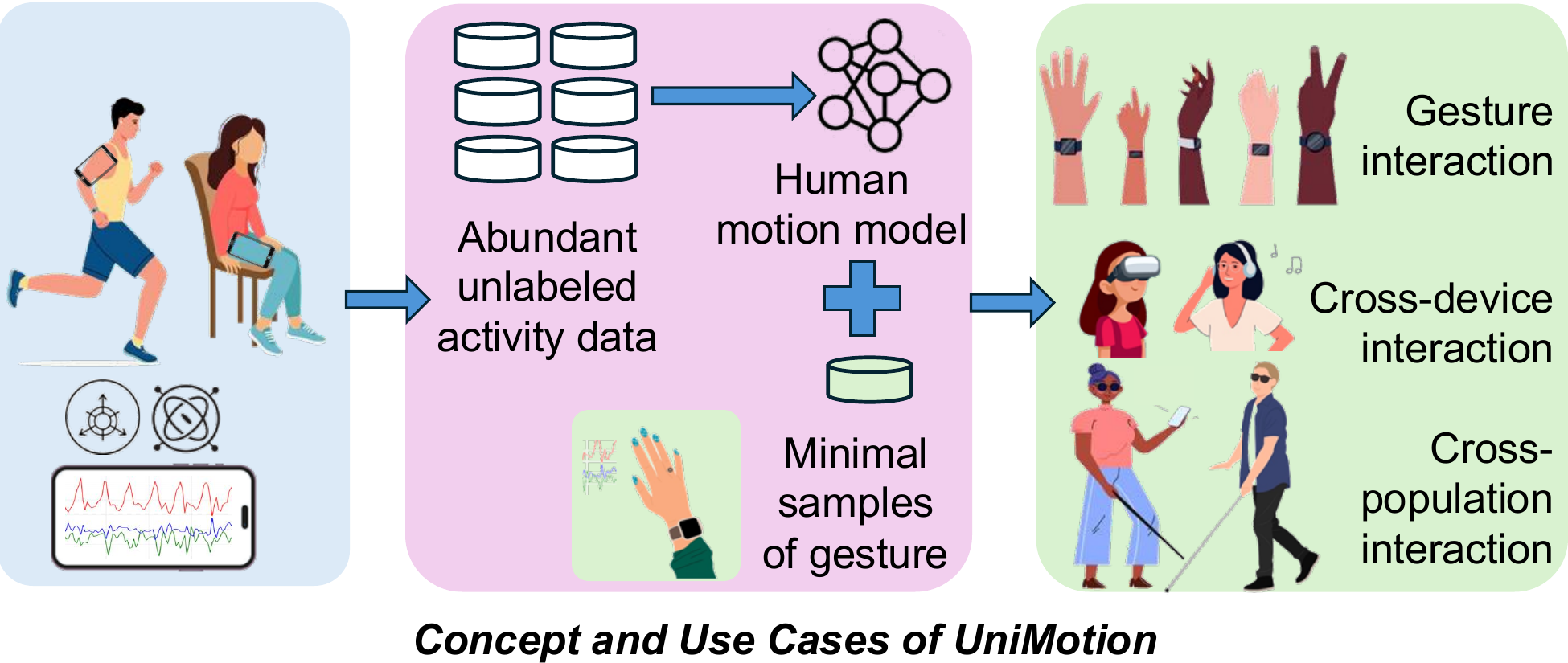}
\caption{\system{}: A framework enabling real-time robust gesture recognition across diverse devices and user populations.}
\label{fig:concept}
%\vspace{-0.5cm}
\end{figure}
IMU-based gesture interfaces are being increasingly adopted as efficient, accessible, and intuitive alternatives to traditional input methods, such as touchscreens and voice. However, current gesture recognition algorithms are tailored to work for specific devices (e.g., smartwatches vs. earbuds) or user populations (e.g., blind vs. sighted users), limiting their generalizability. In this paper, we design \textit{UniMotion}, a generalized IMU-based gesture recognition framework that works across devices and populations with minimal training samples. To overcome the challenges and high cost of collecting large-scale labeled training data, \textit{UniMotion} leverages readily available unlabeled human activity data. The \textit{UniMotion} pipeline comprises two stages: (1) pre-training a motion representation model using abundant unlabeled human activity data, and (2) fine-tuning it with a small amount of labeled gesture data. For pre-training, we introduce a token-based strategy and embeddings that learn to identify and focus attention on the key motion signatures in the temporal data 
For fine-tuning, we design a text-guided classifier that can reliably differentiate between temporally or semantically similar gestures. We evaluate \textit{UniMotion} across both hand gestures (captured through a smartwatch) and earbud gestures (captured through earbuds), using data collected from blind and sighted users. Across these diverse devices and user populations, \textit{UniMotion} achieves an accuracy of 85\%, across an average of 13 gesture classes using only 10\% of labeled data for training. \textit{UniMotion} significantly outperforms state-of-the-art self-supervised learning approaches and specialized gesture recognition models.

\end{abstract}

\maketitle

\section{Introduction} 
\label{sec:intro}

%\aruna{first paragraph is long and rambly, conider making it crisp. Everything you have here is fine, but it is not said in a crisp way}
Gesture interfaces are rapidly gaining popularity for their accessibility, efficiency, and intuitiveness in device interactions~\cite{xu2022enabling, accesswear, serendipity}. Unlike traditional inputs like touchscreens, keyboards, or voice, gestures use natural body movements~\cite{wobbrock2009user, lu2014hand}. These interfaces support a wide range of applications, from subtle mid-air scrolling~\cite{xu2022enabling} and remote appliance control~\cite{freeman2016there} to providing alternate smartphone interaction for users with visual impairments~\cite{accesswear}.
To capture and recognize body movements, researchers have explored various sensing modalities: IMU, cameras, acoustics, WiFi, and EMG. The problem is that modalities other than IMU face several deployment hurdles. Camera-based systems are affected by lighting conditions and occlusion~\cite{chakraborty2018review}, acoustic sensing introduces power and computation costs~\cite{sheng2019wireless}, EMG and WiFi require specialized hardware~\cite{sanchez2020hardware}. 
IMUs, on the other hand, are already built into everyday devices such as phones, watches, and earbuds and can enable real-time gesture recognition on commodity wearables without requiring additional hardware~\cite{honeywell2022imu}. Unlike camera and acoustic-based approaches, IMU sensing is robust to lighting conditions and occlusion~\cite{pan2025diffcap} and incurs low power and computational overhead~\cite{xuan2024review}, making it well-suited for always-on, on-device gesture recognition.
As a result, in this work, we focus on IMU-based gesture recognition. 

There has been extensive research on IMU-based gesture recognition, but current systems require us to rebuild models from scratch for every new device and user demographic~\cite{huang2021tapnet, khanna2024hand, serendipity, xu2020earbuddy, alkiek2023earbender}. These models require collecting extensive labeled data, which is time-consuming and expensive. Further, gesture recognition models trained for smartwatches do not work for earbud gestures, as they are tailored to capture hand movements and have not been trained to capture head movements. Similarly, gesture recognition models trained for sighted users have been shown not work for blind users~\cite{khanna2024hand, accesswear}. %Gestures performed by blind users exhibit different characteristics compared to those of sighted users, requiring researchers to build separate models for each user group. 
Other works have used language supervision for data augmentation, but are computationally expensive and have not been implemented on mobile devices~\cite{miao2024goat, leng2024imugpt}. Therefore, the central question we ask in this work is: 
%\textit{\textbf{Can we build IMU-based gesture recognition models that generalize across devices and populations without requiring extensive labeled data, while enabling real-time performance on commodity devices?}} 

\vspace{2mm}
\colorbox[gray]{0.9}{\parbox{.9\textwidth}{\textbf{Can we build IMU-based gesture recognition models that generalize across devices and populations without requiring extensive labeled data, while enabling real-time performance on commodity devices?}}}

Recent works in Human Activity Recognition (HAR) use self-supervised learning to achieve generalization. In this context, generalization refers to the ability to handle cross-dataset variations, changes in device placement, and domain shifts~\cite{limu-bert, hong2024crosshar, dai2024contrastsense, unihar}. Much like language models~\cite{devlin2019bert, brown2020language} learn from unlabeled text, HAR models learn motion patterns from abundant unlabeled motion sensor data. They operate by masking segments of the motion signal and reconstructing those segments based on the surrounding context. This approach enables high performance in building generalizable HAR models. Once learned from unlabeled data, these models require little labeled data to fine-tune for downstream tasks such as activity classification.
The success of these self-supervised techniques is due to the large amounts of activity data available~\cite{hhar, uci, motion}. Activity data can be passively collected by simply having the user walk/ run around with a phone in their pocket.

%\aruna{what you are saying is that for activity recognition, there has been wokr on generalization using semi-supervised learning, but we want to use it for generalization *and* for gestures where data is not available. I think that point should be made and should be made using simple sentences. What you have here is super clunky. I would say HAR researcher use self-supervised learning for different kinds of generalization, specify what kind of generalization [citations]. But for this, you have a lot of HAR data to train on. We want to use the same idea for generalizing gesture recognition, but we dont have as much gesture data. We want to see if we can use HAR data to train a generalized gesture recognition system. }

We want to apply the same self-supervised learning idea to gesture recognition. However, gesture data, even unlabeled, is not available at scale. Collecting gesture data requires users to repeatedly perform specific swipes or taps, which are not part of natural daily movements like walking. This makes it impractical to gather large datasets. However, we do have an abundance of unlabeled human activity data. 
Since both activities and gestures rely on the same basic movements and are limited by how the body can physically move, such as lifting or rotating a limb, we ask: 

\vspace{2mm}
\colorbox[gray]{0.9}{\parbox{.8\textwidth}{\textbf{Can we use abundant HAR data to build generalized gesture recognition models?}}}
\vspace{2mm}

As a first step, we used existing self-supervised HAR techniques~\cite{limu-bert, unihar, dai2024contrastsense}
%\aruna{only limu bert? dont we do others as well. we should cite all work that we evaluate}
for pre-training and fine-tuned the model for gesture recognition. 
However, our experiments yielded poor gesture recognition accuracy (Section~\S\ref{sec:gesture_recog_performance}). We identify three challenges for why using existing self-supervised techniques {\em as-is} is insufficient: 

(i) Challenge 1: Activities are longer; Gestures are shorter.
%Activities are prolonged and continuous, meaning the information defining the motion is evenly distributed over a long period of time (multiple steps in walking). In contrast, gestures are short and information-dense. The motion-defining part of the gesture is concentrated in a fraction of a second (single tap). 
%Unlike activities, where a self-supervised model can look at the seconds before or after to infer context, gestures lack this extended timeline. If the core part of this short signal is masked, the model has no surrounding context to learn from.
%\aruna{this is ok, but you are introducing masking which has not been mentioned before, consider one more level of abstraction}
Activities are prolonged and continuous, so the motion that defines them is spread across many moments (e.g., multiple steps while walking). Gestures, in contrast, are brief and compact, with the key motion occurring in a very short time span (e.g., a single tap). Because activities last longer, a model has many opportunities to observe how the motion evolves over time. Gestures do not offer this opportunity. If the short, defining part of a gesture is missed, there is little remaining information for the model to learn from.

(ii) Challenge 2: Activities are repetitive; Gestures are isolated. Activities are inherently repetitive. In walking, a step is always surrounded by other steps, creating a predictable cyclic pattern that helps the model learn. Gestures lack this context; a swipe is a single, isolated command. Models that rely on the redundancy found in continuous activities fail when applied to these singular, isolated events.

(iii) Challenge 3: Similar motions, different meanings. Many gestures have similar motion signals but distinct semantics. A \qq{swipe up}  and a \qq{swipe down} share the same speed and energy profile, differing only in trajectory. Similarly, a \qq{tap} and \qq{double tap} share the same motion signal. Standard classifiers struggle to distinguish these fine-grained differences, especially when labeled data is limited.

To address these challenges, we present \system{}, a generalized gesture recognition algorithm that works in two stages: 

\noindent{\bf Stage 1: Token-based pre-training} %We address the first two challenges using a \textbf{token-based pre-training} approach.
As before, we leverage large, unlabeled HAR datasets to pre-train a model in the first stage. Our approach begins by breaking the motion signal into small, meaningful chunks called tokens.
We identify the most critical tokens of the motion signal, which we call the nucleus. The nucleus is the concentrated high-energy segment where the characteristic motion occurs (e.g., foot in-air phase while walking).
We then mask 80\% of the data within the nucleus and train the model to reconstruct these critical tokens. This forces the model to learn even short motion patterns rather than relying on surrounding context. %\aruna{Add a sentence here about how this is different from what happens now. Something like "In contrast, existing works... And then say effectively, this addresses the first two challenges.}
In contrast, random masking used in existing works~\cite{limu-bert, hong2024crosshar, dai2024contrastsense, unihar} often misses the core part of the motion signal and thus fails to work for short gestures. By focusing on the nucleus, \system{} addresses Challenges 1 and 2 by forcing the model to learn from the dense, isolated information that defines the gesture.
Also, we provide additional input signals during this stage to improve the training accuracy. These inputs are information about the nucleus boundaries and which IMU axes contain the most valuable motion data. 

\noindent{\bf Stage 2: Text-Guided Classifier} %In the fine-tuning stage, we fine-tune the models using small amo
%take the pre-trained model and fine-tune for gesture recognition for a specific device or user population using a small amount of labeled data. To build our classifier, we pass target data through the pre-trained model and obtain its learned motion embeddings, which are then fed into the classifier to predict gesture labels. %\aruna{it will be good to make a connection between the embedding and the classification} 
In the second stage, we pass target data through the pre-trained model and obtain its learned motion embeddings, which are then fed into the classifier to predict gesture labels. The classifier learns from a small amount of labeled samples. 
While HAR-based approaches use simple classifiers for the fine-tuning stage, gesture recognition requires better differentiation due to motion similarity between gestures (Challenge 3). % because gestures, like a \qq{tap} and \qq{double-tap}, often result in nearly identical motion embeddings.
Our classifier uses contrastive learning plus a textual description of the gesture to differentiate between similar gestures.
Contrastive learning forces the classifier to push different gestures (negative pairs) apart in the embedding space, and gestures of the same class (positive pairs) close together.
However, traditional contrastive learning treats all negative pairs equally. For example, they would push ``tap'' away from a ``double-tap'' with the same force as from a ``circle''. To overcome this, we use textual description to learn how much to move ``tap'' and ``double-tap'' from the circle gesture, and consequently learn to differentiate between similar-looking gestures. Importantly, the gesture description is only needed for training and does not have to be provided as input at runtime. In \S~\ref{sec:challenges}, we discuss how the contrastive learning approach we introduce is different from existing contrastive approaches used for HAR.

We rigorously evaluate \system{} for gesture recognition accuracy across a diverse range of \textit{public and real-world datasets}, designed to test its \textit{cross-population} and \textit{cross-form-factor} generalization capabilities. Specifically, we consider two distinct target populations (sighted and blind users) and two different device form factors (earbud and smartwatch).~\system{} utilizes a single pre-trained model across all these diverse scenarios, requiring only 10\% labeled data for fine-tuning, yet achieving an average accuracy of 85\%. We conduct extensive comparisons with state-of-the-art works, including ContrastSense~\cite{dai2024contrastsense}, LIMU-BERT~\cite{limu-bert}, and specialized algorithms for blind users~\cite{khanna2024hand} and earable gestures~\cite{alkiek2023earbender}.~\system{} significantly outperforms all of these models by up to 50\% in at least one of the datasets. Also, \system{} is real-time and can run on smartphones, incurring an end-to-end latency of less than 67 msec.

In conclusion, our evaluations definitively show that~\system{} is a real-time IMU-based gesture recognition framework that can adapt effectively across target populations and form factors, addressing limitations of current approaches and paving the way for scalable gesture-based interactions.
%!TEX root = main.tex
\section{Challenges in Generalized Gesture Recognition}
\label{sec:challenges}

\begin{table}[t]
\centering
\small
\renewcommand{\arraystretch}{1.2}
\begin{tabularx}{\linewidth}{|l|X|c|c|c|}
\hline
\textbf{Paradigm} & \textbf{System} & \textbf{Short Motion} & \textbf{Generalization} & \textbf{Real-time} \\ \hline

\multirow{4}{*}{\textbf{Supervised}} 
& TapNet~\cite{huang2021tapnet}, EarBender*~\cite{alkiek2023earbender}, Serendipity*~\cite{serendipity} & \cmark & \xmark & \xmark \\ \cline{2-5}
& Khanna et al.*~\cite{khanna2024hand} & \cmark & \xmark & \cmark \\ \cline{2-5}
& DeepSense*~\cite{yao2017deepsense} & \xmark & \cmark & \cmark \\ \cline{2-5}
& IMUGPT 2.0$^{*}$~\cite{leng2024imugpt}, GOAT~\cite{miao2024goat}, RAG-HAR~\cite{sivaroopan2025rag} & \xmark & \cmark & \xmark \\ \hline

\multirow{2}{*}{\shortstack[l]{\textbf{Masked} \\ \textbf{SSL}}} 
& LIMU-BERT*~\cite{limu-bert}, UniHAR*~\cite{unihar} & \xmark & \cmark & \cmark \\ \cline{2-5}
& CrossHAR~\cite{hong2024crosshar}, SelfPAB~\cite{logacjov2024selfpab} & \xmark & \cmark & \xmark \\ \hline

\multirow{2}{*}{\shortstack[l]{\textbf{Contrastive} \\ \textbf{SSL}}} 
& ContrastSense*~\cite{dai2024contrastsense}, Cosmo~\cite{ouyang2022cosmo} & \xmark & \cmark & \cmark \\ \cline{2-5}
& ColloSSL~\cite{jain2022collossl}, CPCHAR~\cite{haresamudram2021contrastive} & \xmark & \cmark & \xmark \\ \hline

\rowcolor{gray!20}
\multicolumn{2}{|c|}{\textbf{uniMotion}} & \textbf{\cmark} & \textbf{\cmark} & \textbf{\cmark} \\ \hline
\end{tabularx}
\caption{Comparison of prior IMU-based gesture and activity recognition systems along: suitability for short-duration gesture, generalization across devices, and real-time feasibility, highlighting how~\system{} satisfies all three. SSL denotes Self-Supervised Learning. * denotes systems we compare \system{} with in evaluation.}
\vspace{-0.2cm}
\label{tab:related_work}
\end{table}

The goal of \system{} is to build an accurate IMU-based gesture recognition system that generalizes across devices and populations, while enabling real-time performance on commodity devices.
Despite considerable work in human motion recognition (activities and gestures), generalized gesture recognition on resource-constrained devices remains an open problem.
Table~\ref{tab:related_work} shows the related work on human motion recognition that uses IMU sensors (the related work is expanded in \S~\ref{sec:related_work}). 
These works have one or more of the following
limitations—they are not designed to work for short motion durations (our evaluations show that when applied for short motion duration they do not work well); they do not generalize for different tasks; or they do not work in real-time on commodity devices. 
\paragraph{\textbf{Supervised learning methods}: }Gesture recognition works build specialized models for each new device~\cite{huang2021tapnet, alkiek2023earbender, serendipity} or population~\cite{khanna2024hand, accesswear}. 
%\sj{what's the task here? May be merge the next sentence with this one.}. 
%\aruna{I think the first sentence is enough, we can say we will describe them in more detail in related work}
TapNet~\cite{huang2021tapnet} builds a gesture recognition model specifically for smartphones and does not work for different form devices like smartwatches or earbuds. Khanna et al. build a gesture recognition model trained for blind users~\cite{khanna2024hand}. 
This problem of lack of generalization can be solved if we have a lot of training data for gestures (hours of gesture data). However, obtaining training data at scale for gestures is challenging, costly, and time-consuming. %In this context, generalization refers to the ability to handle cross-dataset variations, changes in device placement, device type, and domain shifts.

While recent \textit{language-guided supervised frameworks} such as GOAT~\cite{miao2024goat}, IMUGPT 2.0~\cite{leng2024imugpt}, and RAG-HAR~\cite{sivaroopan2025rag} aim to address the lack of generalization by aligning IMU data with natural language, they do not scale to short-duration gestures. IMUGPT 2.0 acknowledges its inability to generate reliable 3D motion for subtle, fine-grained human movements. GOAT relies on 2-second windows to capture sufficient context for daily activities, which is far too long for millisecond-level gestures. Also, GOAT and RAG-HAR heavily depend on the pre-trained knowledge of the language model, potentially limiting their ability for activity descriptions unseen during pre-training. %Hence, these systems can recognize daily activities such as \qq{walking} or \qq{running}, but they struggle to distinguish the brief gestures. In contrast, \system{} uses textual descriptors that explicitly encode gesture semantics, rather than relying on a language model (Section ~\ref{sec:text_descriptions}).

\paragraph{\textbf{Masking based self-supervised methods}:}
\label{sec:ssl_limitaions}
Researchers in the HAR domain have made progress on this problem of generalization using self-supervised learning techniques~\cite{limu-bert, unihar, hong2024crosshar}.
%\aruna{some of this is repition from intro and can be removed}Obtaining activity data at scale is easier compared to gesture data, as this can be passively collected. 
These techniques use unlabeled human activity data to train a model and fine-tune it for different tasks, and cross-dataset for activity recognition. 
%While these techniques perform well for HAR, they perform poorly when used for gesture recognition.
%The fundamental issue is that existing HAR-based self-supervised methods learn from the surrounding context across activity sequences, while gestures require learning the patterns within brief motion events. 
% 
\begin{figure}[t]
    \centering
    \includegraphics[width=\linewidth]{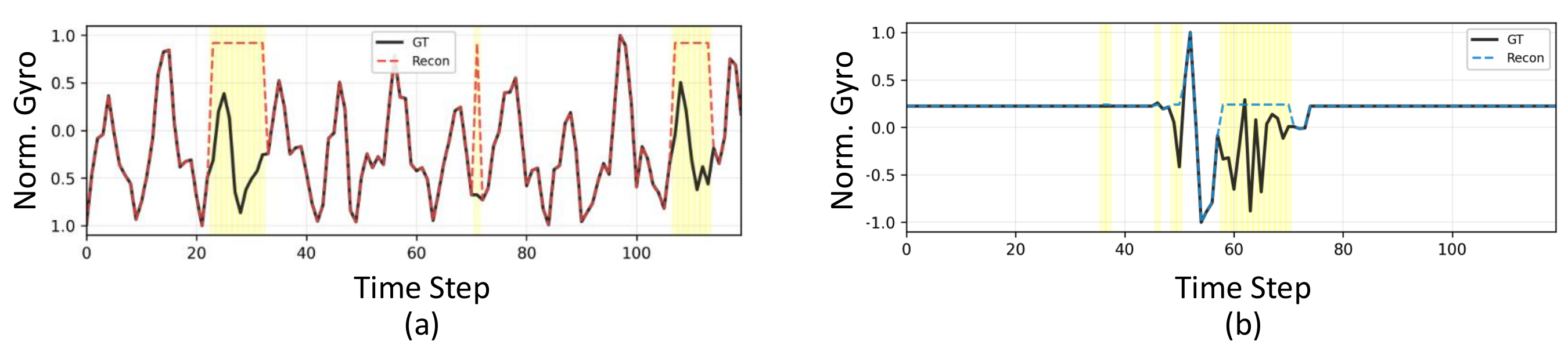}
    \caption{Comparison of random masking~\cite{limu-bert} (yellow shaded regions) on (a) Continuous HAR and (b) Short-duration Gestures. In HAR (a), the model reconstructs the signal by leveraging the surrounding temporal context of repetitive activities. In contrast, for gestures (b), the motion is brief and non-repetitive; random masking fails to capture the fine-grained gesture nuances.}
    \label{fig:reconstruction}
\end{figure}
Training a self-supervised HAR-like model for gesture recognition from scratch is not possible due to the limited availability of gesture data. To illustrate the data scarcity: the smallest publicly available HAR dataset contains 3.5 hours of data~\cite{uci}, whereas the most extensive publicly available gesture dataset is only 0.8 hours~\cite{sony_watch}.

Our goal is to use human activity data for training gesture recognition since both rely on the same human movements, like lifting or rotating a limb. But as discussed in the previous section, using existing techniques~\cite{limu-bert, unihar, hong2024crosshar} for gesture recognition performs poorly. 
%However, we show in our evaluation (\S~\ref{sec:gesture_recog_performance}) that training on human activity data and fine-tuning on gesture data does not work. We evaluate systems like LIMU-BERT and UniHAR and compare them with \system{} to find they perform poorly for gesture recognition.
We find that this poor performance is largely because of how existing HAR-based self-supervised learning techniques perform masked modeling. Current works~\cite{limu-bert, unihar, hong2024crosshar} mask random segments anywhere in the signal, and the model learns to reconstruct them based on the surrounding temporal context. This approach works well for activities because information is distributed relatively evenly throughout repetitive motion patterns.

Figure~\ref{fig:reconstruction} illustrates this. When masking 2 seconds in the middle of a walking sequence (a), the model observes before and after the masked region, identifies the activity context, and predicts continuation of the same pattern. The model learns what typically follows what across the sequence. 
However, for gestures (Figure~\ref{fig:reconstruction}(b)), random masking fails because it lacks the surrounding context to learn from. Unlike human activities, gestures are short and non-repetitive. In gestures, information is concentrated in a brief duration, while the rest of the signal is generic preparation or retraction. If a random mask covers these preparation or retraction areas, the model learns nothing meaningful because those parts carry no important information. Conversely, if the mask covers the gesture-defining region, the surrounding context is again not helpful for reconstruction. This is analogous to masking random characters in BERT instead of whole words; the masking is either irrelevant or destructive to capture the meaningful unit.
%Within self-supervised frameworks, some works use data augmentation~\cite{unihar} to make the models more robust to cross-dataset variations. However, it requires specific domain knowledge to perform augmentation during training.

\paragraph{\textbf{Contrastive-learning based self-supervised methods}:}
\label{para:contrastive_limitations}
Beyond masked modeling, works such as ContrastSense~\cite{dai2024contrastsense} and ColloSSL~\cite{jain2022collossl} leverage contrastive learning to address domain shifts (e.g., cross-device placement) in HAR self-supervised training. These methods use positive and negative pairs to enable domain adaptations, such as cross-location and cross-dataset. In our evaluation, we pre-train ContrastSense on activity data and fine-tune it on gesture data, but this approach performs poorly (\S~\ref{sec:gesture_recog_performance}). This is because such systems apply contrastive learning during the pre-training stage, where class labels are unavailable. Instead, they rely on heuristics of raw motion signal, such as timestamps or sample similarity, to generate positive and negative samples. For instance, ContrastSense~\cite{dai2024contrastsense} defines negative pairs using samples from distant timestamps (high likelihood of different activities) and positive pairs using adjacent samples (high likelihood of the same activity). 

Instead, we find that applying contrastive objectives during the fine-tuning stage is more effective, as class labels are available at this stage (see \S\ref{sec:classifier}). This enables the model to separate gestures based on semantic distinctions rather than relying solely on motion similarity.

\section{Stage 1: Token-based Pre-training} 
\label{sec:pretraining}

\system{} addresses the limitations of existing approaches through a two-stage framework. In this section, we discuss how \system{} uses HAR data to pre-train a model that can be used for gesture recognition. In the next section, we discuss how the model can be fine-tuned for classifying gestures. 
%In Stage 2, we fine-tune the pre-trained model for specific tasks, such as classifying smartwatch gestures.

\begin{figure}[h]
    \centering
    \includegraphics[width=0.7\hsize]{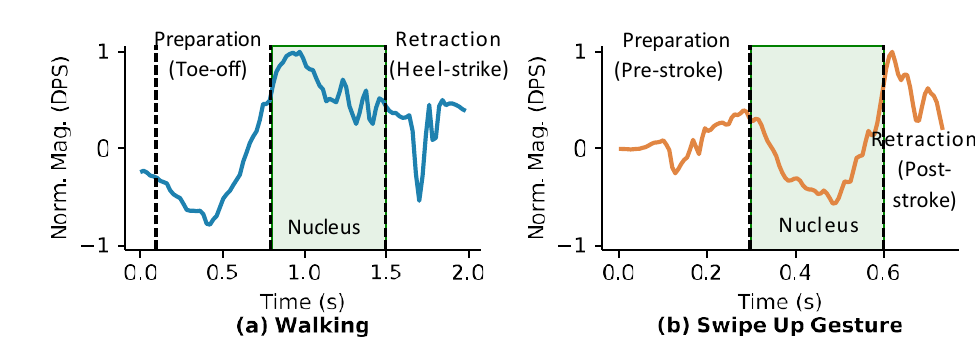}
    \caption{Both (a) walking and (b) swipe gestures exhibit three-phase patterns: preparation, energy-rich nucleus, and retraction. The nucleus contains discriminative information despite differences in duration. }%\aruna{what is Swing? Also, we label pre- and post-stoke for gestures, but why are the labels different for activity?}}
    \label{fig:motion_signature}
\end{figure}

%\aruna{Where are you discussing Figure 3?}
\paragraph{\textbf{Insight}:} 
Gestures are brief motion signals, and the informative part of the gesture that defines it lies in a very short duration of time. To learn short gestures effectively, models must focus on this most informative region. As discussed in \S\ref{sec:challenges}, the problem with existing approaches is that they focus randomly across the motion signal, often missing information from this informative region. 
To learn from this informative region, we leverage the compositional nature of human motion~\cite{yin2014real, benitez2020continuous, accesswear, srivastava2022muteit}. Both gestures and activities are known to be compositional, where the motion can be divided into three phases: the pre-stroke phase (preparation), nucleus (the defining motion), and the post-stroke phase (retraction). Figure~\ref{fig:motion_signature} illustrates this in two signals: (a) a walking activity and (b) a swipe-up gesture. In the case of walking, the actual activity is captured in part when the leg is in the air (swing: from leaving the ground to heel-strike). The portions before and after this can be thought of as getting into position and getting out of position. Similarly, for the \qq{swipe up} gesture, the short 300 msec window shows when the gesture is performed. Pre-stroke is when the user positions their hand to perform the gesture, and post-stroke is when they get out of position. If we can learn the motion signature of this nucleus, then we can perform effective gesture recognition. 
%Therefore, to learn short gestures effectively, models must focus on the most informative region, the nucleus of motion.

%To adapt self-supervised learning for gestures, learning should focus on the most informative part of the signal, where the defining characteristics of a motion occur \aruna{Why? You need to give the reason here again. Is this well known?}Human motion has an inherent compositional structure that we leverage for learning. Both activities and gestures comprise three phases—pre-stroke (preparation), nucleus (the defining movement), and post-stroke (retraction), as shown in Figure~\ref{fig:motion_signature} \aruna{what am I supposed to be seeing in Figure 4?}. Among these phases, the \textbf{nucleus} contains the most discriminative information that distinguishes one gesture from another~\cite{accesswear, khanna2024hand, srivastava2022muteit}. For example, the difference between a swipe up and a swipe down gesture occurs entirely within the nucleus phase~\cite{khanna2024hand} \aruna{based on what?  you are stating is as though it is a fact but without a better explanation of Figure 4 this is not that easy to see}. 

%\aruna{This comes out of nowhere. A bit more preamble here will be useful.}
As discussed previously in \S~\ref{sec:ssl_limitaions}, traditional random masking used in HAR~\cite{limu-bert} does not work when applied to gesture recognition, as it misses the nucleus region. 
Instead, we develop token-based pre-training explicitly targeting the nucleus phase. We tokenize the motion data in three phases and focus on the most important token, the nucleus.
%\paragraph{\textbf{Insight}:} Human motion has an inherent structure that we use for learning. Both activities and brief gestures comprise three phases: pre-stroke (preparation), nucleus (the defining movement), and post-stroke (retraction), as shown in the Figure~\ref{fig:motion_signature}. The nucleus phase contains the most discriminative information that distinguishes one gesture from another~\cite{accesswear, khanna2024hand, srivastava2022muteit}. For example, the difference between a swipe up and a swipe down gesture occurs entirely within this nucleus phase~\cite{khanna2024hand}.  To learn short gestures, we must focus on the most important part: the \textbf{nucleus} of the motion signal. Traditional random masking used in HAR~\cite{limu-bert} ignores this, as we discussed in \S~\ref{sec:ssl_limitaions}. Our token-based pre-training specifically targets the nucleus phase where the defining motion occurs. We tokenize the motion data in three phases and focus on the most important token, the nucleus.\\
\system{} implements this focus and masks in the informative regions to ensure we reconstruct the nuances of a short motion. 
Figure~\ref{fig:reconstruction} shows the pipeline of the pre-training stage. We use unlabeled human activity data to train a self-supervised model that captures motion patterns. We identify the informative parts of the motion and mask within these. We support this learning with other inputs encodings to have a robust model.
%\aruna{this is a jump and does not flow from the previous paragraph, perhaps if the sentence before this is expanded a little more it is flow better}:
To this end, the pre-training stage comprises of the following components:

\begin{figure}[t]
\centering
\includegraphics[width=0.8\linewidth]{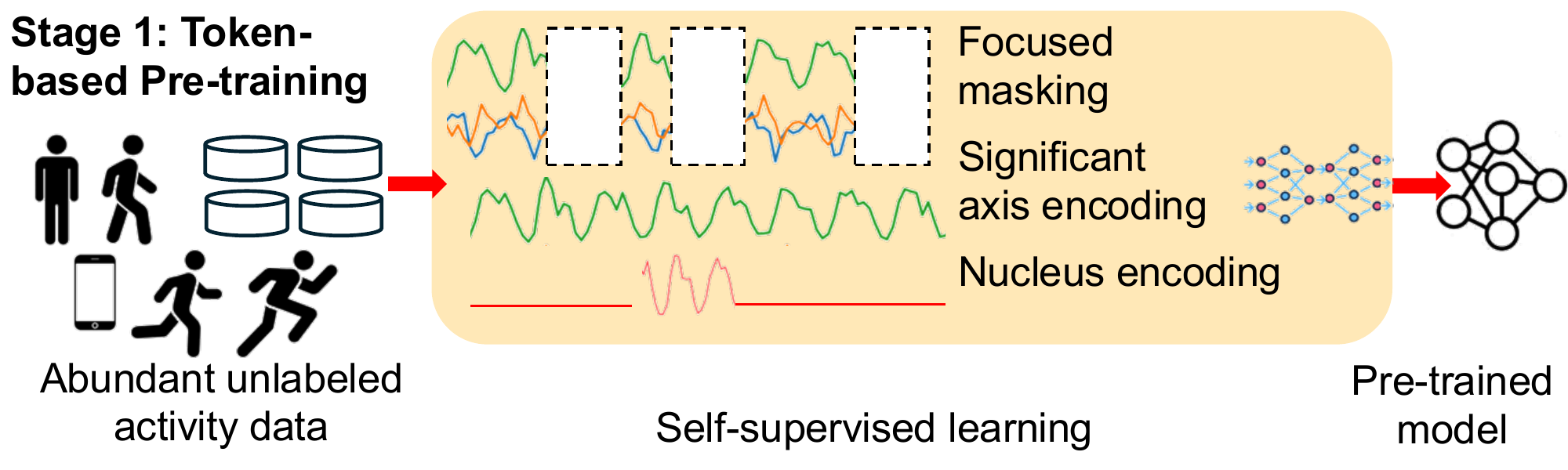}
\caption{Stage 1: Token-based pre-training. The model learns from unlabeled activity data by applying focused masking to the nucleus (high-energy region), while nucleus and significant axis encodings guide attention to discriminative motion patterns.}
\label{fig:pretraining}
\end{figure}

\begin{itemize}
    \item \textbf{Nucleus Identification:} We implement a selection mechanism that automatically detects high-energy motion events in the unlabeled activity data (e.g., the "heel strike" in a walking sequence).
    \item \textbf{Focused Masking:} Unlike random masking as used in HAR approaches~\cite{limu-bert, unihar}, we place 80\% of our masks \textit{inside} the nucleus and only 20\% outside. This forces the model to reconstruct the detailed internal dynamics of the motion event, rather than guessing based on the surrounding context.
    \item \textbf{Input Encodings:} We support this learning with two specific encodings: \textit{Nucleus Encoding}, which explicitly marks the boundaries of the event, and \textit{Significant Axis Encoding}, which identifies the sensor axis containing the relevant motion (e.g., the X-axis for a horizontal swipe), allowing the model to ignore noise.
\end{itemize}
%We use abundant human activity datasets and employ self-supervised learning techniques to train the model. The model learns the patterns \textit{within} a motion event rather than the context \textit{around} it. This representation transfers seamlessly from continuous activities to isolated gestures.

\subsection{Nucleus Identification}
\label{sec:token_definition}
%\aruna{What is shown in figure 4?}As shown in Figure~\ref{fig:motion_signature}, 
Nucleus is the defining phase of a motion event. It is a brief (200--300ms) burst of acceleration that distinguishes a motion event from another. In contrast to Natural Language Processing (NLP), where ``tokens'' (words) are discrete and clearly separated by spaces, IMU data is continuous. There are no explicit boundaries separating a ``nucleus'' from the low-energy preparation or retraction phases.
To enable the nucleus-based tokenization for self-supervised learning, we should automatically be able to detect these regions within continuous, unlabeled activity data. We use the observation that the nucleus is characterized by the high-energy changes in the motion signal~\cite{accesswear, khanna2024hand}.  We compute an energy metric that captures the intensity of motion across all sensor dimensions:
\begin{equation}
energy(t) = \sqrt{\sum_{i \in \{x,y,z\}} (acc_i(t)^2 + gyro_i(t)^2)}
\end{equation}
Using this energy profile, we detect segments where energy significantly changes:

\begin{equation}
\label{eq:nucleus}
    Nucleus = \{t : |energy(t+\Delta t) - energy(t)| > motion\_thresh\}
\end{equation}
where $motion\_thresh$ determines the detection sensitivity.

This process generates a binary mask, $M_{nucleus}$, where $M_{nucleus}(t) = 1$ if time $t$ falls within a nucleus region, and $0$ otherwise. This mask is used for both our input encoding and our masking strategy.

\subsection{Focused Masking}
%HAR-based self-supervised learning approaches~\cite{limu-bert, unihar, hong2024crosshar} mask randomly across the motion data. This random masking approach does not work for gestures as it forces the model to learn form surrounding context, while gestures require learning the patterns within brief motion events (see \S~\ref{sec:ssl_limitaions} why random masking fails for gestures).

To enable learning the patterns of the nucleus, we propose a focused masking strategy. We skew the masking probability distribution heavily towards the nucleus:
\begin{equation}
    P(mask_t) = 
    \begin{cases} 
    0.8 & \text{if } t \in Nucleus \\
    0.2 & \text{if } t \notin Nucleus
    \end{cases}
\end{equation}
We apply masking to 80\% of the samples within the nucleus region and only 20\% to regions outside (to maintain learning diversity). Furthermore, within the nucleus, we mask contiguous spans (4--5 samples) rather than single points.
This configuration changes what the model learns. 
If the model observes an accelerating slope leading to a masked nucleus region, it cannot simply interpolate to the end; it must understand the motion pattern to predict whether the motion continues accelerating or begins decelerating. It learns to reconstruct the motion patterns—the ``shape'' of the force, rather than just the continuity of the line (Figure~\ref{fig:reconstruction}). 
The pre-training objective is to reconstruct the motion data in the masked regions:
\begin{equation}
\mathcal{L}_{pretrain} = \sum_{t \in M} (x_t - \hat{x}_t)^2
\end{equation}
where $M$ is the set of masked indices, dominated by nucleus positions, and $\hat{x}_t$ is the reconstructed output.

\subsection{Input Encodings}
\label{sec:embeddings}
To supplement our focused masking strategy, we provide two additional inputs to the model.
These inputs explicitly mark the nucleus boundaries and identify the IMU signal axis that is most relevant to the motion.  

\begin{figure}[t]
\centering
\includegraphics[width=\linewidth]{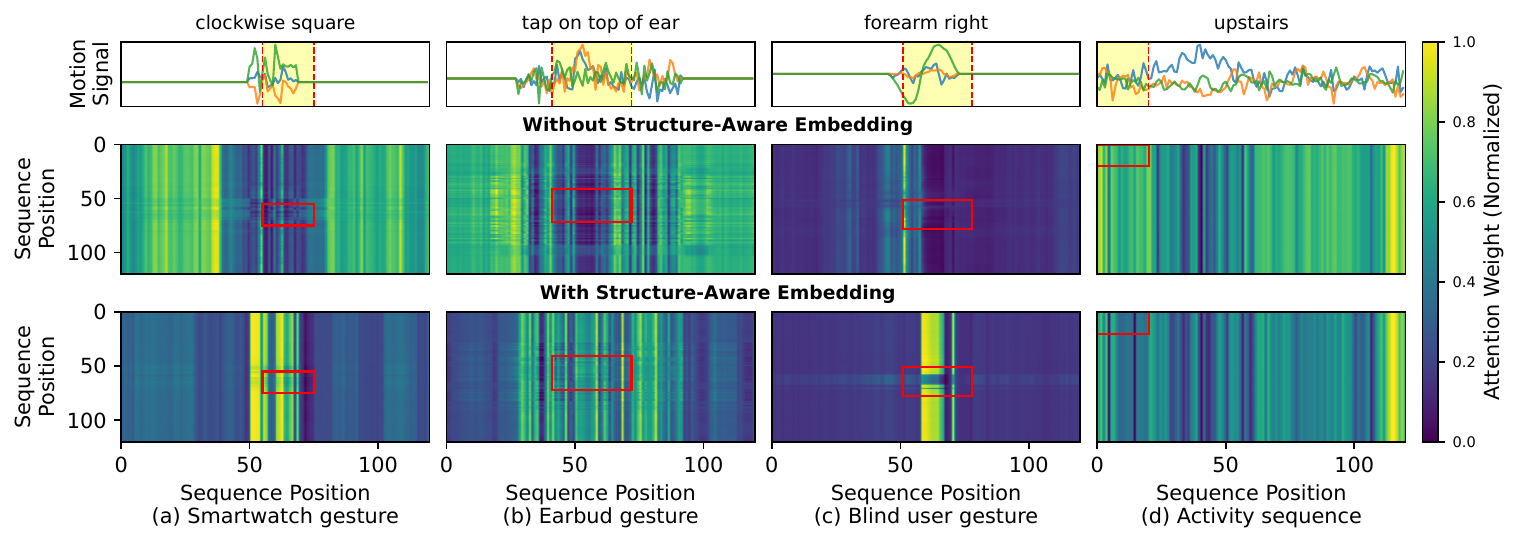}
\caption{Comparison of attention patterns across different motion sequences with and without token-based pre-training. The heatmaps show attention weights between sequence positions, with brighter colors indicating stronger attention. Red rectangles highlight the nucleus regions of each motion activity. Token-based pre-training produces focused vertical attention bands in the nucleus regions; without this guidance, the attention is scattered. On average, token-based pre-training reduced reconstruction MSE by 8.7\% for these sequences.}
\label{fig:attention}
%\vspace{-0.5cm}
\end{figure}

\noindent \textbf{Nucleus Encoding:} This input encoding uses the binary mask derived from Equation~\ref{eq:nucleus} to explicitly mark the boundaries of the nucleus. It signals to the model: ``This is the region where the motion patterns matter the most.'' We map the binary mask to a learnable vector. Time steps falling within the nucleus receive a distinct vector compared to the other time steps. This allows the self-attention mechanism to amplify focus on the useful token and focus less on the preparation/retraction phases. Without this guidance, attention disperses uniformly, missing the critical local patterns (see Figure~\ref{fig:attention}).

\noindent \textbf{Significant Axis Encoding:} Even with guided attention, directional ambiguity remains unsolved. Walking upstairs and downstairs produce identical step patterns except for gravitational axis differences. Similarly, swipe left and swipe right differ only in directional components. We identify the discriminative axis within each token by analyzing rotational patterns across gyroscope dimensions during the motion signature period. The axis with the highest cumulative rotation receives additional attention weight through a binary mask. This ensures tokens encode directional information alongside temporal patterns, creating representations sensitive to both motion dynamics and spatial orientation. Without this component, the model learns directionally invariant features that cannot distinguish opposing movements.

We augment these encoding with a positional encoding. Positional encodings have been used in IMU-based self-supervised methods to provide temporal ordering~\cite{limu-bert}. Transformer models without this positional encoding lack the sense of a temporal sequence. 
The input to the model is a combination of all these encodings:

\begin{equation}
    E = E_{input} + E_{position} + E_{nucleus} + E_{sig\_axis}
\end{equation}

$E_{input}$ represents the linear projection of raw IMU sensor values, and $E_{position}$ provides temporal context. Together, these inputs guide self-attention from uniform temporal processing to token-focused attention.

\subsection{Attention Analysis and Validation} 

We validate that our token-based pre-training successfully creates focused attention through comparative analysis with HAR-based pre-training~\cite{limu-bert}. Figure~\ref{fig:attention} visualizes attention patterns across four motion types: smartwatch gestures, earbud interactions, blind user movements, and climbing activities. Without token-based pre-training, attention disperses uniformly across sequences, treating all temporal positions equally. With focused strategy, attention concentrates into vertical bands precisely aligned with detected nucleus tokens—the model learns to focus where information exists. 
This targeted attention yields concrete improvements: 8.7\% reduction in reconstruction MSE across test sequences. More importantly, this validates our tokenization approach. The model no longer wastes capacity learning from uninformative preparation/ retraction regions but instead develops representations centered on discriminative motion patterns. 
These token-aware representations transfer directly to downstream tasks like fine-tuning for gesture classification.
The model    already understands where to look for critical information, enabling accurate classification even with minimal labeled examples of the target downstream task. 
%\aruna{this is a nice section, I wonder if we expand it a bit more}

\section{Stage 2: Text-guided Classifier}
\label{sec:classifier}

%Stage 1 provides us with a pre-trained model that understands motion patterns from activity data. 

In Stage 2, we use a small amount of labeled data to fine-tune the pre-trained model for gesture classification. %learn a gesture classifier adapt this model for gesture recognition for a diverse set of tasks including gesture recognition on smartwatch, earbuds or different populations. 
%At run time, we pass the target data through the pre-trained model to obtain motion embeddings. In Stage 2, we build a task-specific classifier that can categorize motion embeddings into gesture classes.

\paragraph{\textbf{Insight}:} The pre-trained model captures motion patterns, but it lacks an understanding of gesture semantics. 
While activities in HAR are often distinct, many gestures share nearly identical signals that simple classifiers struggle to separate, especially when gesture data is limited. 
For example, a \textit{tap} and \textit{double-tap} produce similar signal peaks despite representing different actions. However, their semantic difference can guide the classifier to separate them in the embedding space based on not only the motion signal but also their semantics. % spacing them out by their meaning rather than just their motion signal appearance.

%\aruna{This is different from what you say in the intro.Also you need to reiterate (and use a backward pointer) to discuss why this is differnet from related work}
To this end, we use a text-guided contrastive classifier that combines motion patterns and gesture semantics.
%Our approach combines motion embeddings with textual descriptions of each gesture's purpose and characteristics. 
%We use contrastive learning to separate gestures that have similar motions but different semantics. 
%For example, the classifier learns that "tap" and "double-tap" should different despite their similar motion patterns, while gestures with related semantic meanings but different motions can be closer together. 
%\aruna{This next bit could we described in section 2 and then just alluded to over here.}Unlike the contrastive-based pre-training methods discussed in \S~\ref{sec:ssl_limitaions}, we apply contrastive learning during the fine-tuning stage. This allows the system to use class labels and semantic knowledge to separate gestures by actual intent rather than raw signal appearance.\\
\begin{figure}[t]
    \centering
    \includegraphics[width=0.8\linewidth]{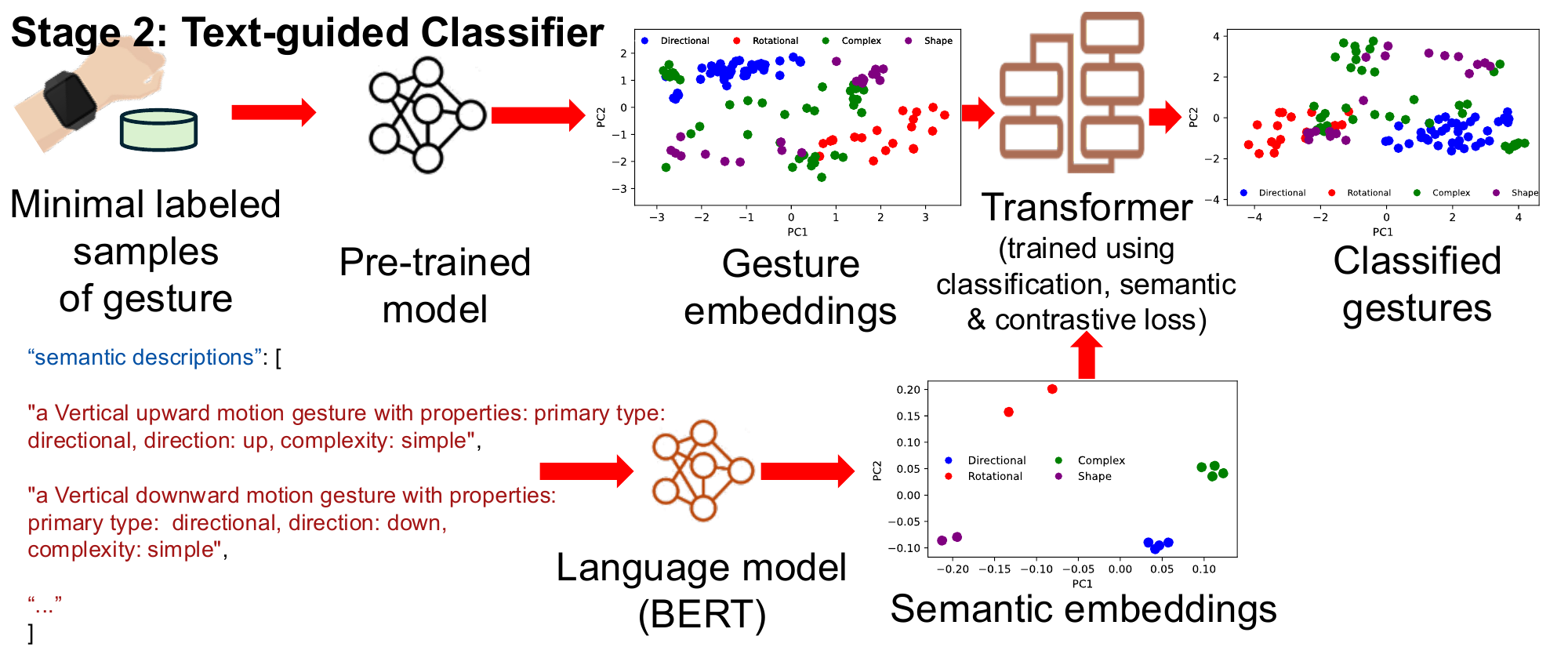}
    \caption{Stage 2: Text-guided Contrastive Classifier. Motion embeddings from the pre-trained model are combined with semantic embeddings derived from gesture descriptions through BERT. The classifier uses semantic and contrastive losses to organize the embedding space, enabling accurate gesture classification with minimal labeled examples.}
    \label{fig:classifier}
\end{figure}
Figure~\ref{fig:classifier} shows the fine-tuning pipeline of \system{}. We pass target data through the pre-trained model to obtain motion embeddings. We then use a small number of labeled examples of the target task to train a classifier that distinguishes between these embeddings.
We build this text-guided classifier in two steps: first, we create text descriptions that capture the semantic properties of each gesture, and second, we use contrastive learning to organize the embedding space based on both motion patterns and semantic meanings. Note that textual description is used only during training; this input is not needed during inference time. We use 10\% of the total labeled data for fine-tuning

\subsection{Text Descriptions}
\label{sec:text_descriptions}
%The dynamic margins and weights used in the contrastive process are derived from text embeddings that encode the semantic properties of each gesture.
In this section we describe the process of generating text embeddings that aid the contrastive learning process. 
Each class in the target dataset is assigned a text description to encode semantic properties using a structured template that specifies the class' category, orientation, and complexity level:
\begin{center}
\begin{minipage}{0.5\textwidth} 
\begin{verbatim}
"a <direction> <type> gesture with properties:
primary type: <category>, direction <orientation>,
complexity: <complexity-level>"
\end{verbatim}
\end{minipage}
\end{center}
These descriptions systematically capture movement direction (up, down, left, right, circular), motion type (swipe, rotation, shape-tracing), higher-level categorization (stationary, directional, rotational, shape, tap), and complexity (simple, complex). For instance, a description for \textit{swipe up} gesture becomes: \begin{verbatim} "an upward swipe gesture with properties: primary type: directional, direction: up, 
complexity: simple." \end{verbatim}
This approach creates semantic relationships between gesture types that mirror human understanding. \textit{Swipe up} and \textit{swipe down} are recognized as related directional gestures, while \textit{circle} and \textit{figure-eight} share characteristics of a shape-tracing gestures. 

We take these text descriptions as input and pass them through a BERT language model. The output is a set of text embeddings that represent each gesture class in a higher-dimensional space. By using these embeddings, the model captures the semantic relationships between gestures. Descriptions with similar meanings are mapped to mathematical points that are closer together in this space.
%\aruna{Who proives this description? How big does it need to be? How do you know it is good enough? How do you provide it at runtime?}
It is important to note that this text description generation is a one-time pre-processing step. The developer manually provides one short description for each class in the target dataset using the structured template discussed. Because there is only one description per class, the total amount of text is very small.
Once these descriptions are processed through BERT to generate the text embeddings, these embeddings remain fixed. They are simply retrieved from storage during the contrastive learning process, meaning the model does not need to re-process text during training and there is no additional computational overhead.

\subsection{Contrastive Learning with Text Guidance}
\label{sec:contrastive_trasformer}
%\aruna{The first 2 paragraph is repetitive, I feel I have be reading the same thing for three sections. }
To distinguish gestures with similar motion patterns, we employ a contrastive learning objective. 
%This process organizes the embedding space by pulling positive pairs (same class) together and pushing negative pairs (different classes) apart. Contrastive learning alone ignores the semantic relationships between gesture types, some gestures should be pushed further apart than others based on semantic meaning.
%Infact, traditional contrastive learning~\cite{dai2024contrastsense} objectives are used in the pre-training stage and they treat all negative pairs equally. They push apart samples from different classes with uniform force because they do not have semantic information during the pre-training stage.
%This is problematic because gestures from different classes often share nearly identical motion signals. Without understanding a gesture's meaning, the system may incorrectly group these similar-looking signals as positive pairs or fail to push them apart as negative pairs.
%We incorporate semantic knowledge into the contrastive learning process. 
The key contribution lies in initializing contrastive distances based on semantic relationships from the text descriptions rather than treating all negative pairs uniformly. When the model encounters negative pairs (samples from different gesture classes), we initialize their separation distance based on the semantic distance between their corresponding text descriptions. This ensures that physically similar but semantically distinct gestures like \textit{swipe up} versus \textit{swipe down} maintain appropriate separation in the embedding space. We describe how we get the text descriptions for each class in the dataset in \S~\ref{sec:text_descriptions}. 

In order to use contrastive objective combined with textual guidance we employ three complementary loss components that work together to create well-separated embeddings:

(1) \textbf{Classification Loss} establishes basic decision boundaries using cross-entropy loss. This is a commonly used loss function in many classification tasks.

%\aruna{Say that this is used in all classification problems}
(2) \textbf{Semantic Loss} enforces appropriate separation distances based on semantic relationships by using a dynamic margin to ensure semantically distinct gestures maintain separation even when motion patterns look similar.
\begin{equation}
\mathcal{L}_{\text{semantic}} = \sum_{i,j} w_{ij} \cdot \max(0, m_{ij} - d(f_i, f_j))
\end{equation}
where $w_{ij}$ is a weight derived from the semantic ambiguity between classes, $m_{ij}$ is a dynamic margin (larger for dissimilar classes and smaller for similar ones), and $d(f_i, f_j)$ is the Euclidean distance between feature vectors.

(3) \textbf{Contrastive Loss} pulls together samples from the same gesture class while pushing apart samples from different classes to maximize similarity within classes and ensure clear separation in the embedding space.
\begin{equation}
\mathcal{L}_{\text{contrastive}} = -\sum_{i} \log \frac{\exp(z_i^T z_{\text{same}} / \tau)}{\sum_{j \neq i} \exp(z_i^T z_j / \tau)}
\end{equation}
where $z_i$ are normalized feature vectors, $z_{\text{same}}$ represents features from the same class as $z_i$, and $\tau$ is a temperature parameter controlling the sharpness of the distribution.

Our final training objective is a weighted sum of these components, with the classification loss serving as the primary objective (implicitly weighted by 1), and semantic and contrastive losses added with gradually increasing weights:
\begin{equation}
\mathcal{L}_{\text{total}} = \mathcal{L}_{\text{classification}} + w_s \cdot \mathcal{L}_{\text{semantic}} + w_c \cdot \mathcal{L}_{\text{contrastive}}
\end{equation}

This dynamic weighting allows the model to first establish basic decision boundaries through classification loss, then progressively incorporate semantic relationships and contrastive separation as training progresses. The result is an embedding space where gestures are organized not just by motion similarity, but by semantic meaning.
Gestures with similar functions cluster together, while semantically distinct gestures maintain clear separation even when their motion patterns overlap. This semantic organization enables accurate classification even with limited labeled examples, as the model can leverage learned semantic relationships to make informed predictions about gesture classes with minimal direct training data.

%!TEX root = main.tex
\section{Implementation}
\label{sec:implementation}

% We implement our system with a transformer-based architecture. 
In this section, we detail our implementation choices, model architecture, and training procedure.
% Our system is implemented using PyTorch. 

\subsection{Model Architecture}
Our system consists of two main components: (1) pre-training a transformer-based model to understand the motion patterns, and (2) learning a text-guided classifier for gesture classification. The input to the model is IMU data, chunked in windows of 120 samples, re-sampled at 20Hz. If the signal length is $<$120, we append zeros at the start and end. This window length is an accepted practice in HAR models~\cite{limu-bert, unihar}.

We normalize the raw input accelerometer and gyroscope data before passing it to the transformer model. The transformer model processes the data through multiple input encodings to generate motion embeddings that capture the discriminative patterns of motion. The model learns to reconstruct masked segments of the input to ensure these motion embeddings contain meaningful representations.
For downstream gesture classification, the motion embeddings are processed through a classifier that uses semantic information to recognize gestures across different form factors and user populations.

\begin{itemize}[leftmargin=*]
\item{\textbf{Pre-training Phase (Stage 1):}}
Our transformer model architecture is inspired by self-supervised works in HAR~\cite{limu-bert, unihar}. We re-imagine this pipeline by integrating domain-specific input encodings (\S~\ref{sec:embeddings}) and replacing random masking with a focused masking strategy.
%\aruna{extend is a weak word, is that all your doing? just adding input encoding?}
The input to the model is 6-axis raw IMU data, projected through a linear layer to form the initial signal representation. This representation is then combined with positional, nucleus, and significant axis input encodings that guide attention to the discriminative motion patterns. The transformer consists of 4 attention heads with a hidden dimension of 72 and a feed-forward dimension of 144. During pre-training, the model learns to reconstruct masked segments of the input sequence through a linear layer followed by layer normalization and GELU activation, focusing 80\% of the masking on nucleus regions as described in \S\ref{sec:token_definition}.

\item{\textbf{Fine-tuning Phase (Stage 2):}}
For gesture classification, we employ a transformer-based classifier that takes the motion embeddings from the pre-trained transformer model as input. The classifier consists of a transformer with 4 attention heads and 2 layers, followed by mean pooling over the sequence length. The transformer applies multi-head self-attention mechanisms and feed-forward processing to identify important patterns in the gesture sequence and generate two outputs: (1) a classification head that predicts gesture labels, and (2) a projection head implemented as a two-layer MLP that creates normalized motion embeddings for contrastive learning. During training, semantic embeddings derived from text descriptions of each gesture class are combined with motion embeddings to organize the embedding space based on both motion similarity and semantic meaning, as described in \S\ref{sec:contrastive_trasformer}. 
%\aruna{do we need a figure here? maybe not but something to think about}
\end{itemize}

\subsection{Training Procedure}
We train our system in two distinct phases.
During pretraining, we use unlabeled IMU data obtained from public human activity datasets (see \S~\ref{sec:datasets}). We implement an focused masking strategy where approximately 15\% of the sequence values are masked (80\% of which are masked in the nucleus regions). The model learns to reconstruct these masked values.
We use the Adam optimizer with a learning rate of 1e-3 and train for 3200 epochs, with early stopping, batch size of 128. The loss function is MSE between the reconstructed and original values. We use 80\% of the activity data for training this model, and use the remaining 20\% for validation and test sets. 

For the gesture recognition task, we implement a combined loss function comprising classification, text-guided, and contrastive loss.
We employ dynamic weights for the text-guided and contrastive loss, which increase gradually during training from 0.1 to their maximum values of 0.3 and 0.5, respectively. This allows the model to first focus on basic classification before incorporating more complex text-guided and contrastive objectives. We train the classifier with a batch size of 64 for 200 epochs using the Adam optimizer with a learning rate of 1e-3. For training the classifier, we use only 10\% of the labeled data, sampled in a stratified manner to ensure a balanced representation across all classes. We selected the above-mentioned hyperparameter using grid search.

%!TEX root = main.tex
\section{Evaluation Setup}
\label{sec:eval_setup}
In this section, we detail the dataset, metrics, and baselines we use to evaluate~\system{}.

\subsection{Datasets}

\begin{table}[h]
\begin{tabular}{|l|c|c|c|c|}
\hline
\textbf{Dataset} & \textbf{Sensors} & \textbf{Classes} & \textbf{Users} & \textbf{Samples} \\
\hline
SU Earbud (Custom)& A,G & 7 & 9 & 315 \\
BU Watch (Custom)& A,G & 12 & 10 & 1200 \\
SU Watch (Publicly available) & A & 20 & 8 & 3200 \\
\hline
HHAR (Publicly available) & A,G & 6 & 9 & 9166 \\
UCI (Publicly available)& A,G & 6 & 30 & 2088 \\
Motion (Publicly available)& A,G & 6 & 24 & 4534 \\
Shoaib (Publicly available)& A,G,M & 7 & 10 & 10500 \\
\hline
\end{tabular}
\caption{Summary of datasets used for evaluating ~\system{}. Top 3: gesture datasets, bottom 4: activity datasets. (A=accelerometer, G=gyroscope, M=magnetometer, SU=Sighted user, BU=Blind user.)}
\label{tab:datasets}
%\vspace{-1cm}
\end{table}

\label{sec:datasets}

To show the performance of \system{} across form factors and populations, we evaluate it on two custom-collected and one public gesture recognition dataset, covering multiple target populations (sighted and blind users) and form factors (earbuds and smartwatches). We also evaluate \system{} on four publicly available HAR datasets for activity recognition. Table~\ref{tab:datasets} summarizes all datasets. Our data collection was conducted following approval from our institute’s Institutional Review Board (IRB). Our dataset selection follows established practices in recent self-supervised and supervised IMU systems for gesture recognition and HAR~\cite{limu-bert, unihar, hong2024crosshar, khanna2024hand, di2025microcontroller} and is designed to evaluate five dimensions of generalization.
\begin{itemize}
    \item The blind user dataset represents a challenging test case, as blind users exhibit higher motion variability than sighted users due to a lack of visual feedback during training~\cite{khanna2024hand}. Successfully recognizing gestures from this population demonstrates the robustness of \system{} to user variance beyond typical inter-user variability.
    \item We evaluate across three distinct form factors—earbuds, smartphones, and two types of smartwatches (Fossil Gen 5, Sony SmartWatch) with different IMU characteristics, sampling rates (10–100 Hz), and wearing positions, showing performance across heterogeneous hardware.
    \item The earbud dataset, with seven gestures and limited labeled samples, evaluates whether our text-guided classification enables learning from minimal data, where collecting extensive labeled datasets is infeasible.
    \item Beyond gesture recognition, we evaluate \system{} on HAR benchmarks to demonstrate its expandability to human motion beyond gestures. This approach achieves HAR performance on par with or better than recent systems, validating that token-based pre-training generalizes across different motion types.
    \item To assess real-time performance, we implement the gesture classifier on a commodity smartphone (Samsung Galaxy S20) and evaluate on-device inference performance, showing that~\system{} can operate under practical latency and resource constraints.
\end{itemize}

While we acknowledge that these datasets remain controlled compared to unconstrained real-world scenarios, they represent the standard evaluation setting in gesture recognition research. Our contribution focuses on moving from random masking to nucleus identification and focused masking, evaluated using benchmarks that enable direct comparison with prior work. In the future, we will explore more diverse real-world deployments. Next, we detail our datasets used for evaluation.
\smallskip

\noindent \textbf{Gesture Recognition Datasets.} 
\begin{itemize}[leftmargin=*,topsep=0pt]

\item \textit{Blind User (BU) Watch Gesture:} We collected this dataset from 10 blind participants (5 male, 5 female) between ages 38-64, with 8 participants blind since birth and 2 participants blind since ages 3 and 11, respectively. Each participant performed 12 different gestures 10 times each while wearing a Fossil Gen 5 smartwatch, and sampled the data at 100 Hz for the accelerometer and gyroscope. These gestures included five forearm directional gestures, five compound gestures, and two shape gestures. Figure~\ref{fig:blind_confusion_matrix} shows the 12 gestures used. We choose these gestures because they have previously been explored for accessible interactions~\cite{accesswear, khanna2024hand}. The duration of the gesture averages 1.3 seconds, which is longer than gestures performed by sighted users. This dataset allows us to evaluate cross-population performance, as the high inter-user variance among blind users demonstrates the system's strong generalization capabilities. 

\item \textit{Sighted User (SU) Watch Gesture:} This dataset~\cite{sony_watch} is publicly available and includes accelerometer data from a Sony SmartWatch worn on the user's right wrist. Eight sighted users performed twenty repetitions of twenty different gestures, totaling 3200 sequences. Gestures were manually segmented by users tapping the smartwatch screen at the beginning and end of each repetition. Data was collected at 10 Hz. 
Using this dataset allows us to evaluate the system's ability to accurately recognize a high number of distinct gesture classes (20), showing our text-guided classifier's capacity to differentiate subtle variations in movement.

\item \textit{Sighted User (SU) Earbud Gesture:} We collected earbud data from 9 sighted participants (3 male, 6 female) between ages 22-29, performing seven gestures on the earbud: tap, double tap, long press, swipe up, swipe down, rotate on the surface, and tap on the lower end. Each gesture was repeated five times. Each gesture lasted about ~0.8 seconds. Similar earable gestures have previously been explored for interaction~\cite{alkiek2022eargest, rateau2022leveraging}. We collected this data using~\cite{earsense}, accelerometer, and a gyroscope sampled at 20 Hz. We use this dataset to evaluate the performance of ~\system{} for different form factors, specifically for earworn devices. This dataset, while featuring fewer gestures (7 classes), is noteworthy for its relatively small number of samples (315), showing the system's performance with limited training data. 

\end{itemize}

\noindent \textbf{Human Activity Recognition Datasets.} We also evaluate \system{} on four public HAR datasets used in prior work~\cite{limu-bert, unihar, hong2024crosshar}. 
\textit{HHAR}~\cite{hhar} contains accelerometer and gyroscope data from 9 users performing 6 activities with 6 phone types worn at the waist, sampled at 100-200 Hz. \textit{UCI}~\cite{uci} includes 30 volunteers performing 6 basic activities with a waist-mounted smartphone at 50 Hz. \textit{MotionSense}~\cite{motion} comprises accelerometer and gyroscope data from 24 diverse participants performing 6 activities with an iPhone in their front pocket at 50 Hz. \textit{Shoaib}~\cite{shoaib2014fusion} collected data from 10 participants performing 7 activities with smartphones at five body positions, recording accelerometer, gyroscope, and magnetometer readings at 50 Hz. 

We use the unlabeled \textit{HHAR} dataset for pre-training our model; 80\% of the data goes in training, 10\% in validation, and the remaining 10\% for testing.
We use \textit{UCI}, \textit{MotionSense}, and \textit{Shoaib} datasets in a cross-dataset manner (train on one and test on another) to evaluate \system{} for the human activity recognition task.

\begin{comment}
\begin{figure}[t]
    \centering
    \includegraphics[width=\linewidth]{figures/output.png}
    %\vspace{-0.6cm}
    \caption{}
    %\label{fig:main_res}
    %\vspace{-0.5cm}
\end{figure}    
\end{comment}

\subsection{Baselines}
We compare ~\system{}'s performance with state-of-the-art models in both supervised and self-supervised learning paradigms. In addition, we also compare the performance of~\system{} against specialized gesture recognition systems that are designed for specific form factors or user populations. 
The selection of these baselines was based on their performance, their status as state-of-the-art (SOTA), and code availability (Table~\ref{tab:related_work}). For example, we selected Khanna et al.~\cite{khanna2024hand} for blind user gesture recognition, while ContrastSense~\cite{dai2024contrastsense} was chosen as the representative for contrastive learning-based systems, as they are the SOTA. Also, IMUGPT 2.0~\cite{leng2024imugpt} is included as the representative for text-based models due to its code availability. %By including these specific SOTA systems, we ensure a fair and direct comparison across all paradigms: supervised, self-supervised, and specialized systems. 
For all the baselines, we perform hyperparameter tuning for epochs, number of layers, and batch size to perform a fair comparison. 

\subsubsection{\textbf{Supervised and Self-supervised baselines}}
\begin{itemize}[leftmargin=*,topsep=0pt]
\item \textbf{DeepSense}~\cite{yao2017deepsense} (supervised). It is a gesture recognition model based on supervised learning. It applies the Fourier Transform to the raw IMU data and feeds the frequency domain features (i.e., magnitude and phase pairs) to the neural network. It adopts a deep learning structure to fuse data from multiple sensors and accordingly extract temporal features.

\item \textbf{LIMU-BERT}~\cite{limu-bert} (self-supervised). LIMU-BERT is an activity recognition model that learns representations from unlabeled IMU data using self-supervised training inspired by the BERT language model. It extracts generalized rather than task-specific features through masked sensor prediction. The learned representations can be used to train downstream task-specific models with limited labeled data (as we do for gesture recognition).

\item \textbf{UniHAR}~\cite{unihar} (self-supervised). UniHAR is an activity recognition model that addresses data heterogeneity in different datasets by augmenting IMU data based on the physics of the sensing process. It integrates federated learning and adversarial training to improve generalization across users. For evaluation, we use the HHAR dataset for fine-tuning UniHAR and deploy it in a data-centralized manner.

\item \textbf{ContrastSense}~\cite{dai2024contrastsense} (self-supervised). ContrastSense is closest to~\system{} and designed for gesture and activity recognition. It tackles domain shifts and label scarcity in wearable sensing through domain-invariant contrastive learning. It leverages unlabeled data with contrastive learning in the pretraining step. We use the token-based pre-training to extract embeddings and use a text-guided classifier for gesture recognition.
\end{itemize}

\subsubsection{\textbf{Gesture Recognition for Specialized Form Factors and Populations}} 

We also compare our generalized gesture recognition system with specialized algorithms:

\begin{itemize}[leftmargin=*]
\item \textbf{EarBender}~\cite{alkiek2023earbender} (supervised). EarBender is a gesture recognition system for touch-based hand-to-ear gestures. It uses IMUs in commodity earables to recognize gestures. The system detects different actions and employs a 1D-CNN for classification and data augmentation. EarBender is designed to be user-invariant. We implemented EarBender's architecture and preprocessing techniques, adapting it to the sighted user's earbud gesture dataset.

\item \textbf{Serendipity}~\cite{serendipity} (supervised). This is a smartwatch-based hand gesture recognition framework. This work trains a personalized SVM classifier. We trained a personalized model for each user in the sighted user hand gesture dataset using the features described in the paper. The features are extracted from the time-series data obtained from both the accelerometer and the gyroscope, and take into consideration the orientation in which the gesture was performed. 

\item \textbf{Hand Gesture Recognition for Blind Users}~\cite{khanna2024hand} (supervised). This work presents a hand gesture recognition system for blind users by 3D gesture trajectory tracking. It employs an ensemble classifier that combines a multi-view CNN model with geometric properties analysis using only gyroscope data. We implemented their ensemble classifier architecture, adapting it to our blind user hand gesture dataset.
\end{itemize}
For both ~\system{} and the baselines, we used 10\% of labeled data for training, sampled in a stratified manner to ensure a balanced representation across all classes: 12 blind user hand gestures, 7 sighted user earbud gestures, and 20 sighted user hand gestures.

\section{Results}
\label{sec:results}
In this section, we present a comprehensive evaluation of ~\system{} for gesture recognition for cross-form factor and cross-population setups, on publicly available and custom-collected datasets. We first summarize our key findings. 
\begin{itemize}[leftmargin=*]
    \item \system{} achieves >85\% accuracy in gesture recognition across two wearable form factors (smartwatch, earbuds) and two user populations (blind, sighted). The second-best-performing baseline performs 50\% worse on at least one dataset. %  outperforming state-of-the-art baselines by 
    \item \system{} acts as a generalized model, outperforming recognition algorithms specially designed for different form factors or user populations. It achieves 85\% earable gestures (surpassing EarBender~\cite{alkiek2023earbender}) and, for blind user gestures (surpassing Khanna et al.~\cite{khanna2024hand}), demonstrating strong generalization with limited labeled data.
    \item \system{} not only achieves high accuracy for the gesture recognition task, but also for the activity recognition task, even though \system{} was not designed for activity recognition. \system{}  achieves a similar accuracy (93\%) to state-of-the-art activity recognition baselines. 
    \item \system{} can efficiently run on smartphones. It only incurs a low end-to-end latency of 67 msec for gesture recognition.
\end{itemize}

\subsection{Gesture Recognition Across Populations and Form Factors}
\label{sec:gesture_recog_performance}

We evaluate~\system{} for: sighted users (SU) performing hand gestures, sighted users performing earbud gestures, and blind users (BU) performing hand gestures. For all three scenarios, we use the same pre-trained model.
\begin{figure}[h]
    \centering
\includegraphics[width=\linewidth]{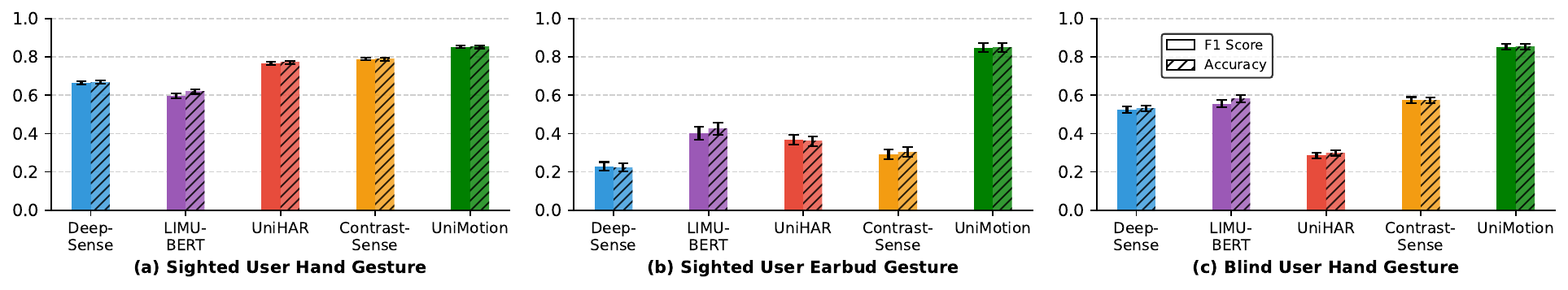}
    \caption{Comparison of F1-score and accuracy of ~\system{} against baseline approaches across three scenarios: (a) Sighted Users Hand Gesture Recognition (20 classes), (b) Sighted Users Earbud Gesture Recognition (7 classes), and (c) Blind Users Hand Gesture Recognition (12 classes). ~\system{} outperforms the baselines.}
    \label{fig:main_res}
\end{figure}

Figure~\ref{fig:main_res} shows the F1-score and accuracy for ~\system{} compared to generalized state-of-the-art supervised and self-supervised models. 
For \textit{Sighted user-hand gesture recognition} (Figure~\ref{fig:main_res}a), ~\system{} achieves >85\% accuracy and F1-score, significantly outperforming DeepSense (67\%) and LIMU-BERT (62\%). UniHAR and ContrastSense show accuracy close to \system{} at 79\% and 80\% respectively. Both these systems use a domain adaptation technique to learn a gesture classifier. Another reason for the good performance of these two models is that the hand gesture dataset is relatively large compared to the other two datasets. This gives enough samples, even with 10\% to adapt their pre-trained models during fine-tuning. We also evaluated IMUGPT 2.0~\cite{leng2024imugpt} for smartwatch gesture recognition; however, it consistently underperformed all generalization-focused baselines (41\% accuracy). As a result, we omit IMUGPT 2.0 from subsequent cross-form-factor and cross-population evaluations.

For the \textit{Sighted user-earbud gesture recognition} dataset, ~\system{} maintains 85\% accuracy while all baseline accuracies, including UniHAR and ContrastSense, drop (DeepSense: 22\%, LIMU-BERT: 42\%, UniHAR: 39\%, ContastSense: 31\%). There are two main reasons for this relatively low baseline performance. First, with respect to DeepSense, the earbud dataset is much smaller in size compared to the other two datasets, making supervised learning more challenging. With respect to LIMU-BERT, UniHAR, and ContastSense, the earbud gestures are short-lived and subtle. This makes learning much harder for these models because they perform random masking as opposed to our focused masking. \system{} is able to overcome both the limited dataset and the short gesture duration to achieve high accuracy. 
Figure~\ref{fig:sighted_earbud_confusion_matrix} shows the confusion matrix for sighted users' earbuds gestures- more pronounced gestures like tap and double tap have higher accuracy than finer gestures of swipe up and swipe down. We attribute~\system{}'s distinguishing ability to our text-guided classifier. 

On the blind user-hand gesture recognition dataset, ~\system{} achieves 85\% accuracy, outperforming all baselines. ContractSense is the second-best-performing baseline with an accuracy of 59\%. All baselines perform better in this dataset compared to the earbud dataset because the gestures are longer~\cite{khanna2024hand} and more data is available for fine-tuning. 
%\aruna{One additional point to make here is that the second-best performing keeps changing as well.} 
It is also worth noting that no single baseline consistently takes the second-best spot. The top-performing baseline changes depending on the specific dataset or task.
However, \system{} is the only model that works with high accuracy for diverse form factors and target populations.

\subsection{Comparing with Specialized Gesture Recognition Algorithms}
\begin{figure}
    \centering
    \includegraphics[width=0.6\linewidth]{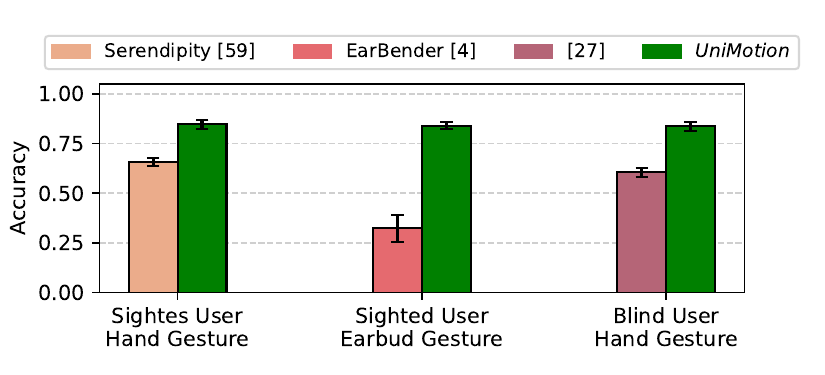}
    \vspace{-0.3cm}
    \caption{Comparison of accuracy of ~\system{} across three baselines for sighted user hand and earbud (SU-HG, SU-EG) gesture and blind user hand gesture (BU-HG) with recent wearable gesture recognition systems.}
    \label{fig:main_res_new}
    \vspace{-0.2cm}
    \end{figure}
\begin{figure}
    \centering
    \includegraphics[width=0.6\hsize]{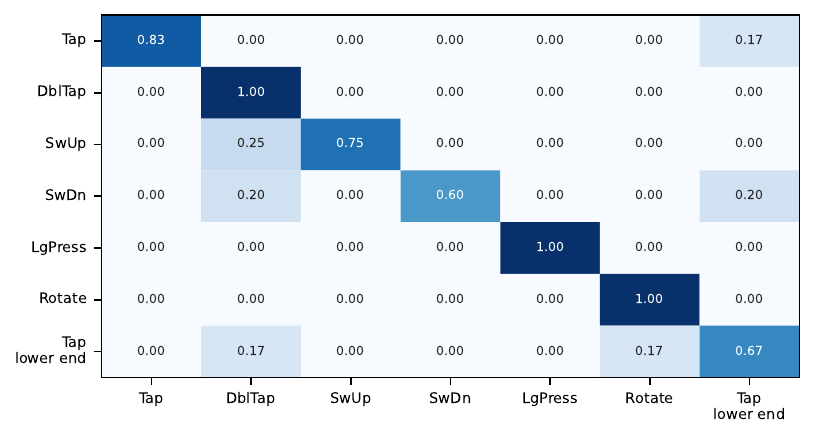}
    \vspace{-0.5cm}
    \caption{Confusion matrix for sighted users' earbud gestures. Pronounced gestures like tap and double have slightly higher performance than more stubble gestures like swipe up and swipe down.}
    %\vspace{-0.5cm}
\label{fig:sighted_earbud_confusion_matrix}
\end{figure}
\begin{figure}
    \centering
    \includegraphics[width=0.6\hsize]{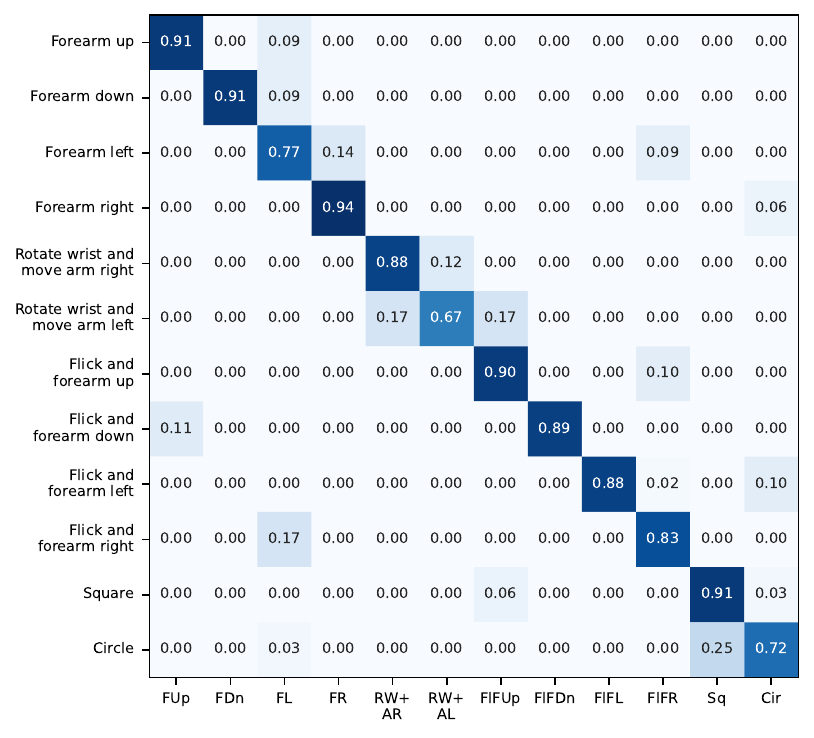}
    \caption{Confusion matrix for blind users' hand gestures. Simple directional movements achieve >90\% accuracy, while compound gestures show minimal confusion between similar movements.}
    \label{fig:blind_confusion_matrix}

\end{figure}

After comparing with generalizable recognition models, we evaluate \system{} against gesture recognition algorithms specifically designed for individual datasets: Serendipity~\cite{wen2016serendipity} for hand gestures from sighted users, EarBender~\cite{alkiek2023earbender} for ear-based gestures, and~\cite{khanna2024hand} for hand gestures from blind users. The goal is to assess whether these specialized models can outperform \system{} on the datasets they were designed for under limited data conditions. We use 10\% labeled data to train all models. Figure~\ref{fig:main_res_new} shows that none of the dataset-specific models achieve the accuracy of \system{}. As before, this is because the baseline models rely on relatively large labeled datasets for effective training. In contrast, \system{} leverages unlabeled data through token-based pretraining to learn meaningful human motion representations, enabling the text-guided classifier to build accurate classifiers from only a few labeled examples per class. This is further reflected in the confusion matrices (Figures~\ref{fig:sighted_earbud_confusion_matrix} and~\ref{fig:blind_confusion_matrix}), where \system{} exhibits reduced confusion between gestures with similar motion. For instance, Figure~\ref{fig:sighted_earbud_confusion_matrix} shows clear separation between \qq{tap} and \qq{double tap}, while Figure~\ref{fig:blind_confusion_matrix} shows high discrimination between \qq{forearm up} and \qq{forearm down}.

\subsection{Effect of Increasing Gesture Classes}
\begin{figure}[h]
    \centering
    \includegraphics[width=0.5\linewidth]{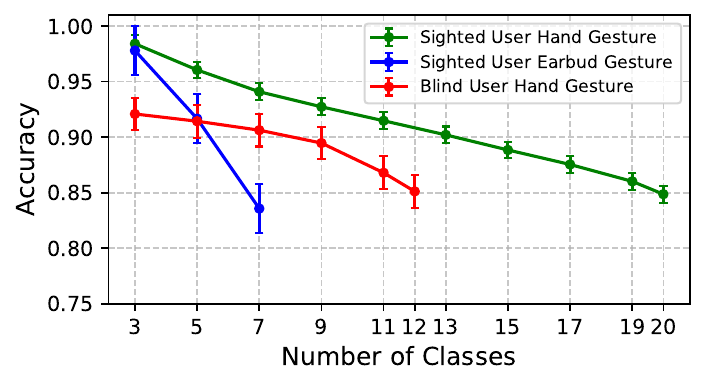}
    \caption{Performance of ~\system{} as the number of gesture classes increase. \system{}'s performance gracefully degrades as the number of classes increases, but the overall accuracy is still high even when disambiguating across a large number of classes. }
    \label{fig:num_classes}

\end{figure}

As interaction systems evolve, users often require larger gesture vocabularies—whether to support more complex interactions as they become familiar with a system or to provide comprehensive accessibility options for diverse user needs. We evaluate ~\system{}'s ability to handle these requirements by increasing the number of classes and reporting classification accuracy in Figure~\ref{fig:num_classes}. For \textit{SU-Hand gestures} (20 classes), ~\system{} achieves above 85\% accuracy even at 20 classes, with a gradual decrease from approximately 98\% at 3 classes. This slow degradation curve suggests scalability for typical hand gesture interactions. Similarly, for \textit{SU-Earbud gestures} (7 classes), ~\system{} maintains high accuracy (above 95\% for 3-5 classes), though with a more pronounced drop to about 84\% at 7 classes. This steeper decline reflects the inherent challenge of distinguishing between subtle earbud gestures as the discrimination task becomes more complex. For\textit{ BU-Hand gestures} (12 classes), ~\system{} achieves over 90\% accuracy at 3 classes and maintains performance above 85\% up to 12 classes. These results show ~\system{}'s scalability with an increasing number of gesture classes across different form factors and user populations.

\subsection{Performance with Limited Training Data}
\begin{figure*}[h]
    \centering
    \includegraphics[width=\linewidth]{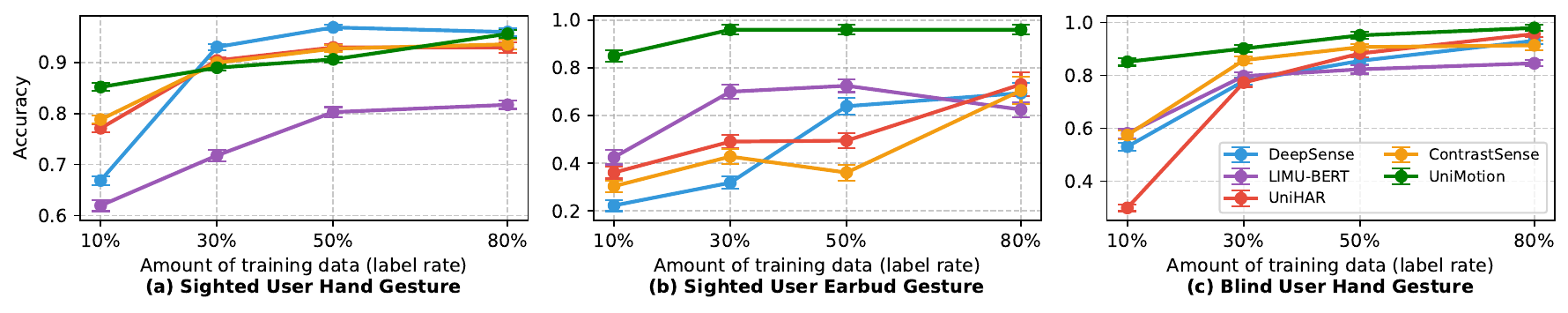}
    \vspace{-0.5cm}
    \caption{Effect of different fractions of labeled training data used to train the gesture classifier. ~\system{} maintains consistently high performance even with limited training data, while baseline models show greater dependence on training data quantity.}
    \label{fig:label_rate}
    %\vspace{-0.5cm}
\end{figure*}

We investigate the performance of~\system{} with varying amounts of labeled training data, ranging from 10\% to 80\% of the available data. Recall that the default is to use 10\% labeled data. This scenario reflects real-world conditions where collecting large labeled datasets can be challenging, especially for novel gesture types or diverse user populations. For \textit{}{SU-Hand gesture recognition} (Figure~\ref{fig:label_rate}a), ~\system{} achieves high accuracy (~85\%) even with only 10\% labeled data and maintains consistent performance as the label rate increases. It's worth noting that because 10\% of SU-Hand gesture data is approximately 10× and 2.6× larger than other datasets, there is still sufficient data for other baselines to perform reasonably well with hand gestures, which might not be the case for newer classes of gestures like earbud movements or data from accessory populations.  

For \textit{SU-Earbud gesture recognition} (Figure~\ref{fig:label_rate}(b)),~\system{} maintains >80\% accuracy for all label rates, showing its ability to extract meaningful features even from minimal data. The baselines show more significant performance variations, with DeepSense, in particular, struggling at low label rates (35\% accuracy at 30\% label rate). This highlights one of the key differentiating factors of our system from baselines: utilizing supportive embeddings and text-guided contrastive learning to learn and classify meaningful representations. Similarly, for \textit{BU-Hand gesture recognition} (Figure~\ref{fig:label_rate}c), ~\system{} performs better across all label rates. These results demonstrate ~\system{}'s capability to work with limited labeled data, which is valuable in contexts where labeled data collection is challenging or expensive. 

\subsection{Ablation study}
\label{sec:ablation_study}
We conduct an ablation study and report accuracy in Table~\ref{tab:ablation}, showing the impact of removing individual components on system accuracy. Specifically, we (1) replace focused  masking with random masking during pretraining, (2) omit nucleus and significant-axis encoding during pre-training, (3) remove textual descriptors during the contrastive classification, (4) omit contrastive classification, and (5) replace the transformer-based classifier with a lightweight MLP. Our results show that removing any of the components leads to at least a 15\% drop in accuracy, highlighting the importance of each component to ~\system{}'s performance. 

Interestingly, removing \textit{Motion Signature and Signal Axis Embeddings} causes significant accuracy drops across all three evaluation scenarios, demonstrating their critical importance for cross-form factor and cross-population gesture recognition tasks. 
On the other hand, removing the semantic contrastive classification leads to a substantial accuracy drop for earbud gestures (the middle accuracy value drops from 0.81 to 0.57). This is because earbud gestures involve subtle movements such as swipes up/down that produce very similar signals, making gesture differentiation crucial. 
Finally, the transformer architecture we use for classification has minimal impact on performance, suggesting that \system\ can potentially work well with other classification architectures.

\begin{comment}
\begin{table}[t]
\begin{tabular}{|c|c|c|c|c|c|}
\hline
\textbf{ME} & \textbf{SAE} & \textbf{SE} & \textbf{CL} & \textbf{TC} & \textbf{Acc} \\
\hline
\ding{55} & \checkmark & \checkmark & \checkmark & \checkmark & 0.76, 0.58, 0.77 \\ 
\hline
\checkmark & \ding{55} & \checkmark & \checkmark & \checkmark & 0.76, 0.68, 0.77 \\ 
\hline
\checkmark & \checkmark & \ding{55} & \checkmark & \checkmark & 0.83, 0.57, 0.84 \\ 
\hline
\checkmark & \checkmark & \checkmark & \ding{55} & \checkmark & 0.83, 0.65, 0.73 \\ 
\hline
\checkmark & \checkmark & \checkmark & \checkmark & \ding{55} & 0.82, 0.81, 0.83 \\ 
\hline
\end{tabular}
\caption{Ablation study results when individual system components are removed. Accuracy column: Hand Gesture (SU), Earbud Gesture (SU), Hand Gesture (BU). ME: Motion Signature Embedding, SAE: Significant Axis E, SE: Semantic E, CL: Contrastive Learning, and TC: Transformer-based Classifier.}
\label{tab:ablation}
%\vspace{-1cm}
\end{table}    
\end{comment}

\begin{table}[t]
\centering
\small
\begin{tabularx}{\textwidth}{|>{\centering\arraybackslash}X|>{\centering\arraybackslash}X|>{\centering\arraybackslash}X|>{\centering\arraybackslash}X|>{\centering\arraybackslash}X|>{\centering\arraybackslash}X|}
\hline
\textbf{Focused Masking} & \textbf{Input Encodings} & \textbf{Text Description} & \textbf{Text-guided Contrastive Classifier} & \textbf{Transformer-based Classifier} & \textbf{Accuracy} \\
\hline
\ding{55} & \checkmark & \checkmark & \checkmark & \checkmark & 0.76, 0.58, 0.77 \\ \hline
\checkmark & \ding{55} & \checkmark & \checkmark & \checkmark & 0.76, 0.68, 0.77 \\ \hline
\checkmark & \checkmark & \ding{55} & \checkmark & \checkmark & 0.83, 0.57, 0.84 \\ \hline
\checkmark & \checkmark & \checkmark & \ding{55} & \checkmark & 0.83, 0.65, 0.73 \\ \hline
\checkmark & \checkmark & \checkmark & \checkmark & \ding{55} & 0.82, 0.81, 0.83 \\ \hline
\end{tabularx}
\caption{Ablation study results detailing the impact of individual framework components. Accuracy column: Hand Gesture (SU), Earbud Gesture (SU), Hand Gesture (BU).}
\label{tab:ablation}
\end{table}

\begin{comment}
\noindent \textbf{User Authentication:} We also evaluate the performance of ~\system{} for user authentication based on gesture patterns. Our system successfully identifies users based on their unique gait patterns, achieving high authentication accuracy. This demonstrates ~\system{}'s ability to capture subtle personal characteristics in gesture execution, making it suitable for biometric authentication applications.    
\end{comment}

\subsection{Beyond Gesture Recognition: Cross-task HAR}

\begin{figure*}[h]
    \centering
    \includegraphics[width=\linewidth]{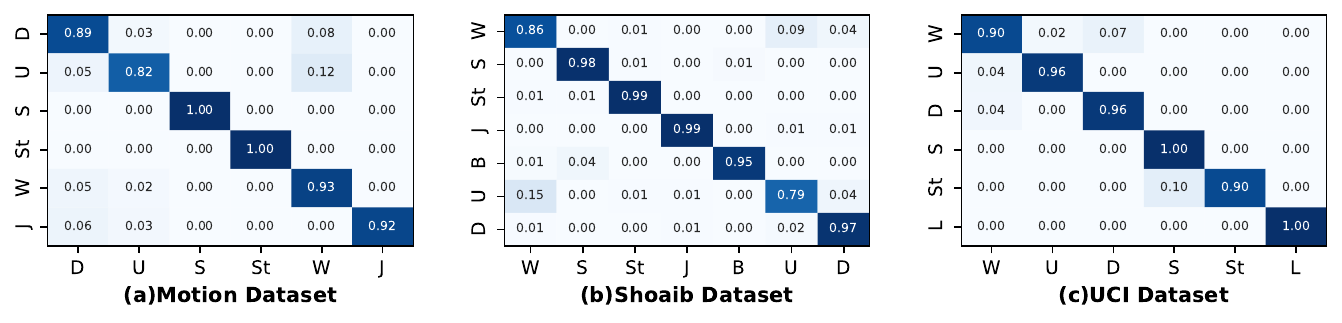}
    \vspace{-0.5cm}
    \caption{Performance of ~\system{} for cross-task from gesture recognition to human activity recognition. The model is pre-trained on the unlabeled HHAR~\cite{hhar} dataset and fine-tuned using 10\% labeled data from the test set. The average accuracy is 93\% across three test HAR datasets: Shoaib, MotionSense, and UCI. }
    \label{fig:conf_acivity}

\end{figure*}

Even though \system{} is designed for gesture recognition, we show that it works well for other related task: human activity recognition (HAR). To evaluate \system{} for HAR, we pre-train the motion model using the HHAR dataset~\cite{hhar} (without any labels) and then test it on the other three HAR datasets after fine-tuning the model with 10\% labeled data from the test set. This is the cross-dataset splits for HAR performed in previous works~\cite{unihar, hong2024crosshar}. Figure~\ref{fig:conf_acivity} shows the confusion matrix when~\system{} is trained on HHAR and tested over the Shoaib~\cite{shoaib2014fusion}, MotionSense~\cite{motion}, and UCI~\cite{uci}. Across these datasets, \system{} achieves an average accuracy of 93\%, even when trained on only 10\% labeled data. This is very close to the state-of-the-art model UniHar~\cite{unihar}, which is specifically designed for HAR (94\% accuracy). UniHar is specifically designed to work well for cross-dataset activity recognition. In contrast, LIMU-BERT~\cite{limu-bert}, which is not designed for this cross-dataset scenario, only achieves an accuracy of 40\% when training on 10\% labeled data. 
The result shows that the token-based pre-training we design in \system{} is able to capture human motion well enough to recognize both short-duration gestures as well as longer activities.

The practical implication is that developers could potentially leverage ~\system{} as a foundation for both gesture interfaces and activity monitoring applications—from smart home systems that recognize fine-grained gestures to healthcare applications that track coarse-grained activities, all with minimal dataset-specific training.

%We evaluate ~\system{}'s cross-task and cross-dataset generalization capabilities by performing HAR. We evaluated the performance of ~\system{} in activity recognition by pre-training on the HHAR dataset and testing on Shoaib, MotionSense, and UCI datasets, with a label rate of low 10\%. As shown in Figure~\ref{fig:conf_acivity}, our approach achieves greater than 93\% accuracy for cross-dataset performance, which is comparable to the SOTA cross-dataset HAR system, UniHAR (94\%). 

\subsection{Smartphone Implementation}
We implement the entire \system{} pipeline on an Android phone and smartwatch. For mobile deployment, we transform our two-stage architecture into a unified TorchScript model, which provides an optimized, serialized representation that can execute efficiently on mobile devices. %We combine the pre-trained motion embedder and semantic contrastive classifier into a single forward pass. 
We utilize PyTorch Mobile's~\cite{PyTorch} 8-bit dynamic quantization, which converts 32-bit floating-point weights to 8-bit integers during inference. This quantization process is applied post-training and optimizes transformer components without requiring model retraining. 
The mobile implementation preserves the motion signature detection and significant axis identification algorithms, executing them efficiently in native Java code before model inference. The quantized model size was $863.16kB$. The end-to-end latency for gesture recognition across 10 trials on a Samsung Galaxy S20 phone was 66.3 msec, well below the latency that can be perceived by humans~\cite{laming1968information}. On average, it took 46 msec to stream gesture data via Bluetooth from the smartwatch to the phone, and it took 20.3 msec to run the model inference on the phone. %\aruna{This last sentence can be made stronger}
Together, these results show that \system{} not only achieves high accuracy for multiple user populations and devices with minimal labeled data, but also supports end-to-end, real-time gesture recognition on commodity smartphones.
%The median latency for running the model inference was 20 msec, with a mean of 20.3 msec.

%!TEX root = main.tex
\section{Related Work}
\label{sec:related_work}
\noindent Gesture recognition systems have evolved from early approaches with hand-crafted features to advanced machine learning techniques extracting higher-dimensional features. Current research in this field primarily takes two directions: traditional recognition methods requiring large labeled datasets, or self-supervised learning approaches that utilize unlabeled data to learn representations, thereby reducing dependency on extensive labeled datasets.

\noindent\textbf{Self-supervised Learning for IMU-based Sensing:} LIMU-BERT~\cite{limu-bert} and its follow-up UniHar~\cite{unihar} are amongst the first papers that use self-supervised learning using unlabeled HAR data; but the focus in both papers is human activity recognition (HAR). We already discussed and demonstrated the shortcomings of these models for the gesture recognition task. CrossHAR~\cite{hong2024crosshar} addresses data heterogeneity challenges through physics-based data augmentation and hierarchical contrastive sensor data pretraining, improving cross-dataset generalization. 
However, CrossHAR has poorer performance compared to UniHar for HAR, and the source code is unavailable, so we omit it from our baseline. The most recent work in this series, ContrastSense~\cite{dai2024contrastsense}, is closely related to~\system{}. It proposes domain-invariant contrastive learning by leveraging domain labels to minimize discrepancies across users and devices and targets both activity recognition and gesture recognition. In contrast to~\system{}, ContrastSense utilizes sample similarity to create negative pairs with the expectation that they are distinct. Since various gestures share comparable motion signals, the system identifies these similar motions as negative pairs and attempts to separate them within the feature space.
Beyond self-supervised learning, transfer learning and domain adaptation techniques have been explored. While some transfer learning systems for gesture recognition require extensive base model training data~\cite{qi2023resource} or report low accuracy~\cite{kang2021wrist}, domain adaptation specifically targets generalization challenges across varying IMU data conditions. Notable works include Soleimani and Nazerfard's~\cite{soleimani2021cross} use of GANs for cross-subject transfer learning in HAR, Sanabria and Ye's~\cite{sanabria2021unsupervised} unsupervised adaptation for heterogeneous datasets, and Hu et al.'s~\cite{hu2023swl} SWL-Adapt for cross-user variability in wearable HAR. Though primarily focused on activity recognition, these methods' strategies for robust generalization are highly relevant to gesture recognition across diverse populations and form factors.

\noindent \textbf{Gesture Recognition for blind users:} Gestures offer significant benefits for accessibility, especially for visually impaired users~\cite{leporini2012interacting, xu2019darkreader, azenkot2013exploring, kane2011usable}. Research shows gestures performed by blind users differ substantially from sighted users, with higher inter-user variance, possibly due to lack of visual feedback, rendering recent recognition works impractical without large datasets~\cite{accesswear}. While Modzelewski et al.\cite{modzelewski2012hand} demonstrate gesture recognition for blind users, their approach requires custom hardware. AccessWear\cite{accesswear} works with commercial smartwatches but supports only 5 gestures. Khanna et al.~\cite{khanna2024hand} expand the gesture set by designing a gesture recognition algorithm that uses 3D representation. Our evaluations show that this technique does not work well when only a small amount of labeled data is available. %but achieve low accuracy with limited data. In contrast, our approach supports a larger gesture vocabulary and achieves higher performance with limited data (see Section~\ref{sec:results}).

\begin{comment}
\noindent\textbf{Deep Learning for IMU-based Gesture Recognition:} Deep learning techniques for IMU-based gesture recognition use accelerometer, gyroscope, and magnetometer data to capture motion patterns~\cite{chen2024deep, jiang2021emerging, noh2024decade, hashi2024systematic}. While early systems relied on signal processing techniques like Kalman filters~\cite{zhou20132d, severin2020head, chan_fusion}, Hidden Markov Models~\cite{jiang2021emerging, wang2023hmm}, and Dynamic Time Warping~\cite{luo2021wearable, liu2023ultrasonic, yang2019imu}, later works used CNNs and RNNs~\cite{mohanty2017deep, chung2019efficient, al2019hand, prakash2020accurate} for extracting temporal and spatial features from IMU streams across various devices~\cite{shin2016dynamic, zhu2018control}. Recent approaches have achieved accuracy exceeding 95\% for predefined gestures but struggle with cross-user performance~\cite{makaussov2020low}. While few-shot learning methods enable personalized gestures~\cite{enabling-hand-gesture, bigdelou2012adaptive}, they still require substantial labeled data (>500 users) to train base models. This dependence on labeled data and poor generalization capabilities across diverse devices and user populations limit the widespread adoption of gesture-based input interfaces.    
\end{comment}

\noindent\textbf{Deep Learning using large datasets:} Several gesture recognition algorithms train deep learning models such as CNNs and RNNs~\cite{mohanty2017deep, chung2019efficient, al2019hand, prakash2020accurate} using labeled data. These systems extract temporal and spatial features from IMU streams, showing promising results for smartphones, smartwatches, and earables~\cite{shin2016dynamic, zhu2018control}. Other systems have utilized commercial wearable devices for gesture recognition. TapNet~\cite{huang2021tapnet}, Serendipity~\cite{wen2016serendipity}, and EarBender~\cite{alkiek2023earbender} are designed to recognize smartphone, smartwatch, and earbud gestures, respectively, using IMU. Previous works have shown that these existing approaches, designed primarily for sighted user gestures, do not work well when applied to blind user gestures~\cite{khanna2024hand}. Our evaluation of Serendipity~\cite{wen2016serendipity}, and EarBender~\cite{alkiek2023earbender} shows that these techniques do not work well even for the scenario they were designed for, with limited data. Recent advancements leverage natural language processing and generative AI, such as IMUGPT 2.0~\cite{leng2024imugpt} and GOAT~\cite{miao2024goat}, to generate synthetic IMU data or align activity representations for improved HAR generalization with minimal data. While promising for data-scarce HAR, they can not be directly applicable to gesture recognition, with its shorter durations and subtle variations, which warrants further investigation.

\noindent\textbf{Alternative sensor modalities for gesture recognition} There has been considerable work on using alternate sensor modalities, beyond IMUs for gesture recognition. For instance, systems have used EMG~\cite{chen2020high, wei2019surface}, microphones~\cite{das2017ultrasound, watanabe2018improving, wang2019hand}, and cameras~\cite{oudah2020hand, murthy2009review, shahi2024vision} amongst other sensors. We start with IMU-based gesture recognition because IMUs are available in most commercial devices. We leave the exploration of alternate sensor modalities for future work. %have been explored for gesture recognition, they face challenges including limited availability in commercial devices, high power consumption, and privacy concerns, which hinder their wide adoption. %These limitations make IMU sensors a more practical candidate for gesture recognition systems that can be deployed across diverse commercial devices.}

%However, these systems do not scale well for different populations due to user-specific features~\cite{khanna2024hand} and show poor performance when tested with a limited amount of data (Section~\ref{sec:results}). To address these shortcomings recent deep learning works have explored few-shot learning. While these systems enable personalized gestures~\cite{enabling-hand-gesture, bigdelou2012adaptive}, they still require substantial labeled data (>500 users) to train base models. This dependence on labeled data and poor generalization capabilities across diverse devices and user populations limit the widespread adoption of gesture-based input interfaces.

In summary, existing gesture recognition approaches require substantial labeled data or struggle with cross-user and cross-device generalization. %, and perform poorly for diverse populations with limited samples. 
Self-supervised learning offers a promising direction but current techniques fail to address the unique characteristics of gesture data—shorter durations, non-repetitive movements, and gesture similarity. %Our work addresses these limitations. %through a motion signature-based framework that specifically learns information-rich segments of motion data while capturing semantic understanding of movements, enabling robust gesture recognition across diverse users and devices with minimal labeled data.

%!TEX root = main.tex

\section{Discussion and Limitations}

\noindent \textbf{Beyond Earbud and Blind User Gesture Recognition:} While~\system{} demonstrates robust performance with earbuds and smartwatches, we envision expanding the system to emerging wearable form factors. Smart rings provide continuous finger and hand motion tracking with minimal user burden, while VR headsets and AR glasses enable head-mounted sensing. Beyond device diversity, we plan to evaluate~\system{} with populations with distinct motor characteristics, particularly individuals with Parkinson's disease whose tremor patterns and movement dynamics vary significantly across users. Such populations may benefit from gesture-based interfaces that accommodate motor variability. Additionally, our current evaluation focuses on indoor environments with minimal ambient noise. In the future, we will evaluate~\system{}'s robustness under real-world deployment scenarios, including noisy environments, moving vehicles, and outdoor settings.

\noindent \textbf{Exploring Gestures across Modalities:} Our framework currently focuses on IMU data, offering advantages in privacy, power consumption, and device availability. In the future, we will extend~\system{} to extract human motion representations from other modalities such as surface electromyography (sEMG) and camera-based vision systems. Rather than combining multiple modality streams, our approach would leverage a unified motion model that can process various input modalities independently, maintaining the same underlying representation of human motion regardless of the sensing technology used.

\noindent \textbf{Transformer Architectures for Time-Series Signals:} While~\system{} demonstrates strong performance for gesture recognition, transformers may not be optimal for continuous time-series signals despite our embedding of the motion signature~\cite{zeng2023transformers, chen2023contiformer}. In future work, we plan to explore specialized architectures like Conformers that combine convolutional networks with transformers to better model local patterns while maintaining global context. These architectural methods could further improve disambiguation between subtly different gestures with similar motion patterns.

%\noindent \textbf{Towards a Foundational Model for Human Motion:} Our North Star goal is developing a unified foundational model for human motion extending beyond recognition tasks. Similar to large language models in NLP, we envision a motion foundation model capable of understanding, generating, and predicting human movements. Such a model could perform trajectory generation, predict biometric signals from motion patterns, or enable novel interaction modalities through motion synthesis, enabling applications in healthcare, accessibility, and human-computer interaction.
\section{Conclusion}
In this paper, we introduce~\system{}, a generalized learning framework for human gesture recognition that works effectively across form factors and user populations. We leverage a two-stage approach: token-based pre-training on unlabeled activity data and a text-guided classifier using contrastive learning with limited labeled gesture data. Our token-based pre-training helps focus attention on the most discriminative part of the motion during pre-training, while our text-guided classifier differentiates between similar gestures when performing classification. We evaluate across different devices: smartwatch and earbud, different populations: sighted and blind users, and different tasks: activity and gesture recognition.~\system{} achieves 85\% accuracy across diverse datasets using only 10\% of labeled data and significantly outperforms state-of-the-art gesture recognition and self-supervised learning systems. Our unified real-time framework eliminates the need for device-specific or population-specific gesture recognition models, enabling more scalable and accessible gesture interfaces.

\bibliographystyle{ACM-Reference-Format}
\bibliography{references} 
\clearpage
\end{document}